\documentclass{emulateapj} 
\nocite{*}

\shorttitle{Isolated OB Associations}

\begin{document} 

\title{Isolated OB Associations in Stripped H~I Gas Clouds}

\author{J. K.\ Werk\altaffilmark{1},
M. E.\ Putman\altaffilmark{1},
G. R.\ Meurer\altaffilmark{2},
M. S.\ Oey\altaffilmark{1},
E. V.\ Ryan-Weber\altaffilmark{3},
R. C.\ Kennicutt, Jr.\altaffilmark{3}, and
K. C.\ Freeman\altaffilmark{4}}

\altaffiltext{1}{Department of Astronomy, University of Michigan, 500 Church St., 
		Ann Arbor, MI 48109, $jwerk@umich.edu$}
\altaffiltext{2}{Department of Physics and Astronomy, The Johns Hopkins University, 
		Baltimore, MD 21218-2686}
\altaffiltext{3}{Institute of Astronomy, University of Cambridge, Madingley Road, Cambridge, 
		CB3 0HA, UK}
\altaffiltext{4}{Research School of Astronomy and Astrophysics (RSAA), 
		Australian National University, Cotter Road, Weston Creek, ACT 2611, Australia}

\begin{abstract}

{\it{HST}} ACS/HRC images in UV (F250W), V (F555W), and I (F814W) resolve three isolated OB associations that lie up to 30 kpc from the stellar disk of the S0 galaxy NGC 1533.  Previous narrow-band H$\alpha$ imaging and optical spectroscopy showed these objects as unresolved intergalactic H~II regions having H$\alpha$ luminosities consistent with single early-type O stars. These young stars lie in stripped H~I gas with column densities ranging from 1.5 - 2.5 $\times10^{20}$ cm$^{-2}$ and velocity dispersions near 30 km s$^{-1}$. Using the HST broadband colors and magnitudes along with previously-determined H$\alpha$ luminosities,  we place limits on the masses and ages of each association, considering the importance of stochastic effects for faint (M$_{V}$ $>-8$) stellar populations.  The upper limits to their stellar masses range from 600 M$_{\odot}$ to 7000 M$_{\odot}$, and ages range from 2 - 6 Myrs.   This analysis includes an updated calculation of the conversion factor between the ionizing luminosity and the total number of main sequence O stars contained within an H~II region. The photometric properties and sizes of the isolated associations and other objects in the HRC fields are consistent with those of Galactic stellar associations, open clusters and/or single O and B stars. We interpret the age-size sequence of associations and clustered field objects as an indication that these isolated associations are most likely rapidly dispersing. Furthermore, we consider the possibility that these isolated associations represent the first generation of stars in the H~I ring surrounding NGC 1533. This work suggests star formation in the unique environment of a galaxy's outermost gaseous regions proceeds similarly to that within the Galactic disk and that star formation in tidal debris may be responsible for building up a younger halo component. 

\end{abstract}
\keywords{galaxies: star clusters -- H~II regions -- intergalactic medium -- galaxies: halos
-- stars: formation}

\section{Introduction}
\label{intro}

H~II regions in gaseous tidal arms, far beyond the rotating disks of galaxies, present an opportunity to study the star formation process in conditions vastly different from those in typical galactic spiral arms. They may even contribute to our understanding of satellite galaxy formation and evolution, halo stellar populations, and the enrichment of the intergalactic medium.  Such intergalactic H~II regions have been found in galaxy clusters, compact groups, and in the halos of galaxies (e.g. \nocite{boquien07, walter,emma,oosterloo,cortese,sakai,gerhard,arnaboldi, mendes04, gallagher01} Boquien et al. 2007; Walter et al. 2006; Ryan-Weber et al. 2004; Oosterloo et al. 2004; Mendes de Oliveira et al. 2004; Cortese et al.  2003; Sakai et al. 2002; Gerhard et al. H~II
2002;  Arnaboldi et al. 2002; Gallagher et al. 2001). In each case, they
represent young stellar associations forming ostensibly where no stars existed
previously. Intergalactic H~II regions  are
distinct from the H~II regions in the outer arms of spiral galaxies
(e.g., \nocite{ferguson98} Ferguson et al. 1998) in that they are further from the
host galaxy, they do not show significant continuum emission, and thus have much higher H$\alpha$ equivalent widths \citep{emma,gerhard,mendes04}.  These regions may represent H$\alpha$-emitting counterparts to the extended UV-disk population of star clusters revealed by GALEX, as they lie at similar distances from the host galaxies, and exhibit similar levels of star-formation \citep{Thilker07}. Much speculation exists as to whether intergalactic H~II regions represent isolated stars, what role tidal interactions play in their formation, whether their stellar populations are typical of young star clusters in quiescent systems, and whether they contribute significantly to the enrichment of the IGM.

\begin{figure*}
\begin{center}
\plotone {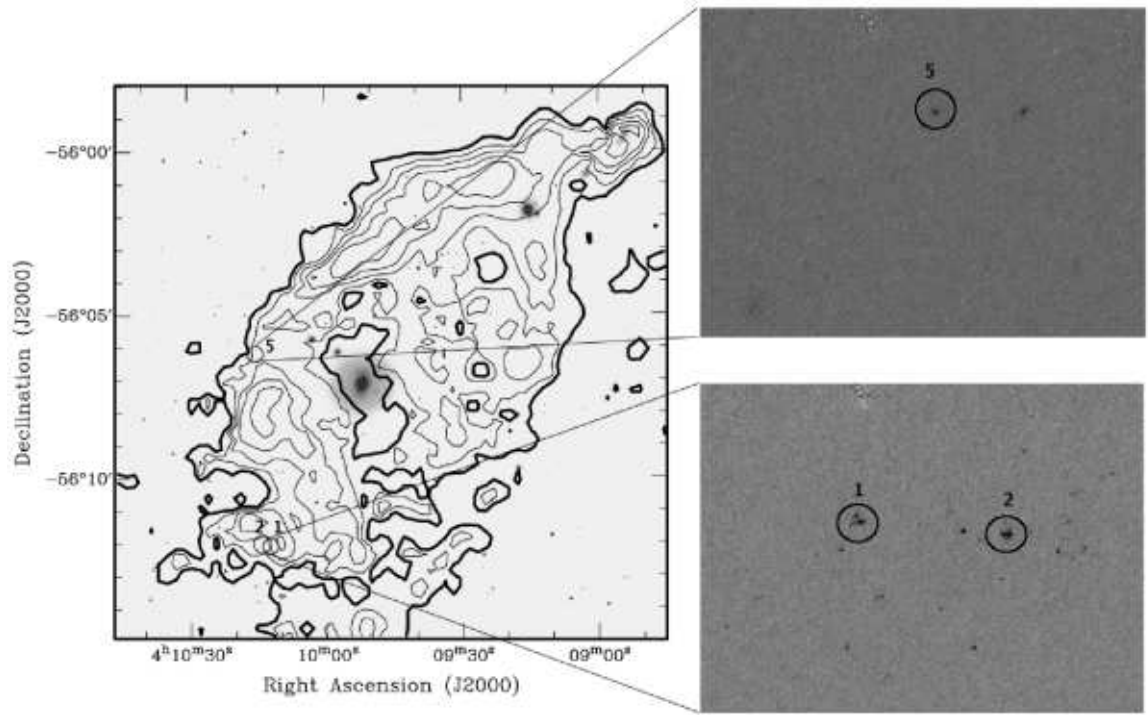}
\figcaption {The H$\alpha$ SINGG image of NGC 1533 overlaid with Australia Telescope Compact Array (ATCA) {\ion{H}{1}} contours and marked with the locations of the individual spectroscopically confirmed H~II regions studied here. The detection images  are shown to the right of the SINGG image. The ATCA {\ion{H}{1}} contours are 1.0 (in bold; 3$\sigma$), 1.5, 2.0, 2.5, and 3.0 $\times 10^{20}$ cm$^{-2}$ and have a resolution of $\sim 1$ arcminute. There is no detectable H~I emission on the optical core of NGC 1533.}
\end{center}
\end{figure*}

Five intergalactic H~II regions, three of which have been spectroscopically confirmed, lie in the vicinity of the galaxy NGC 1533. \cite{emma} discovered these H~II regions in the continuum-subtracted narrow-band image of NGC 1533 taken by the Survey for Ionization in Neutral Gas Galaxies (SINGG, \nocite{SINGG} Meurer et al. 2006). The peculiar ring-like distribution of {\ion{H}{1}} around NGC 1533 as seen in Figure 1 (and figure 2 of Ryan-Weber et al. 2004 \nocite{emma} with higher H~I cut-off levels) may be the result of an interaction with the nearby dwarf galaxy seen in the NW corner, IC2038 (with the SINGG identification J0409-56:S2). No obvious optical counterpart to the disturbed {\ion{H}{1}} appears in either the DSS image or the SINGG image. The velocities of the three spectroscopically confirmed regions coincide well with the {\ion{H}{1}} gas, which is bound to NGC 1533 \citep{emma}. The projected radius from the center of NGC 1533 to the isolated H~II regions ranges from 19 kpc (region 5) to 31 kpc (regions 1 and 2), up to four times the R-band 25$^{th}$ magnitude isophotal radius of NGC 1533. Table 1 summarizes the properties of the three H~II regions presented in \cite{emma} and imaged here. It gives the regions'  positions, projected radii from NGC 1533, velocities determined from their H$\alpha$ emission lines, H$\alpha$ luminosities, and  H~I columns at each region's position. Note that the values in our Table 1 differ from those presented in \cite{emma} because of the recent revision in the measurement of the distance to NGC 1533 using images taken in parallel to those discussed here (see Barber DeGraaff et al. 2007). In this paper, we present {\it{Hubble Space Telescope}} Advanced Camera for Surveys/ High Resolution Channel (ACS/HRC) images and photometry of the three spectroscopically-confirmed isolated H~II regions associated with NGC 1533. These observations enable us to resolve the star-forming regions, identify their ionizing sources, and carry out an in-depth study of their stellar populations.  

\begin{deluxetable*}{ccccccc}
  \tabletypesize{\small}
  \tablewidth{0pc}
  \tablecaption{Properties of the Confirmed Isolated H~II Regions in NGC 1533 \tablenotemark{a}}
  \tablehead{\colhead{H~II}& \colhead{RA}& \colhead{Dec}& \colhead{Separation}&\colhead{Velocity}&
                        \colhead{Log L$_{H\alpha}$ \tablenotemark{b}}&\colhead{N$_{H~I}$}\\
   		 \colhead{Region}&\colhead{}&\colhead{}& \colhead{kpc}&\colhead{km s$^{-1}$}& \colhead{ergs $s^{-1}$}&\colhead{$10^{20}$cm$^{-2}$} }
  \startdata
  
   1&          04 10 13.7&     -56 11 37.5 &    31&   $846\pm50$  &$37.67\pm0.26$& 2.4\\
   2&          04 10 14.5&     -56 11 35.9 &    31&   $831\pm50$  &$37.50\pm0.26$&2.4\\
  5&           04 10 15.7&     -56 06 16.2 &    19&   $901\pm50$  &$37.35\pm0.26$&1.5\\

  \enddata
  \tablenotetext{a} {\scriptsize Summary of properties originally presented in Ryan-Weber et al. (2004): Positions are given in J2000 coordinates, with RA in hours, minutes, and seconds, and Declination in degrees, arcminutes, and arcseconds. Separation is the projected separation in kiloparsecs to the optical center of NGC 1533. The H~I column densities are measured from ATCA maps, at the position of each H~II region. NGC 1533 has a heliocentric velocity measured by HIPASS to be 785 km s$^{-1}$.  The numbers for projected separation and H$\alpha$ luminosity differ from those presented in \cite{emma} because we use the most recent measurement of the distance to NGC 1533, 19.4 Mpc (Blakeslee, private communication).}
  \tablenotetext{b}{\scriptsize As measured in the SINGG H$\alpha$ images}
\end{deluxetable*}

In combination with the HST data, we make use of an ATCA H~I synthesis map of NGC 1533 \citep{emma03} to help constrain some of the details of the formation and evolution of the isolated H~II regions.  Intergalactic H~II regions are often found forming in relatively low column density H~I gas.  
There is no H~I detected to a limit of 2  $\times$10$^{19}$ cm$^{-2}$ at the location of the H~II region near NGC 4388 (\nocite{hioosterloo}Oosterloo \& van Gorkom, 2005), and the H~I surface density in the vicinity of the three confirmed systems, regions numbered 1, 2, and 5 in Figure 1 is $ 1.5-2.5 \times 10^{20}$ cm$^{-2}$ \citep{emma}. Furthermore, the H~I peaks in the neutral gas ring surrounding NGC 1533 and in the stripped material around NGC 4388 do not exhibit any detectable level of star formation.   These observations offer further support for the lack of correlation between star formation rate densities and H~I surface densities on local scales found by \cite{kennicutt07}, unlike what is seen on global galactic scales.  The overall kinematics of the gas in the region of the intergalactic H~II regions may provide clues as to the trigger of the star formation and the potential growth of the stellar associations into a tidal dwarf galaxy.

Independent of the gaseous properties of cluster environments, star-cluster ``infant mortality" suggests that the isolated, young star clusters presented here may be short-lived \citep{rz05,fcw,cfw}. Will supernova and mass loss in these clusters remove enough interstellar matter to leave the stars freely expanding, resulting in the clusters' rapid dissolution? Studies of the Antennae galaxies \citep{fcw,cfw} and the Small Magellanic Cloud \citep{rz05} purport that over half of clusters formed dissolve less than 10 Myr after their formation independent of their initial masses. However, a recent study by \cite{gieles07} concludes that there is little or no ``infant mortality" for ages beyond 10-20 Myrs in the Small Magellanic Cloud, and \cite{degrijs07} find less than 30\% of young ($<20$ Myrs) SMC clusters undergo ``infant mortality." Generally, studies of this kind require a statistical sample of clusters. Instead, resolving a few star clusters in an isolated environment in a period during which they may be actively dissolving offers a different approach to examining their fates.

The photometric properties of the underlying stars in these intergalactic H~II regions can help to answer some of the questions surrounding their formation and evolution. Previous studies have used stellar population synthesis models and evolutionary tracks to infer cluster ages and masses. For example, \cite{whitmore99} identify 3 distinct populations of star clusters in and around NGC 4038/4039 (Antennae Galaxy) using their integrated photometric properties. \cite{gerhard} determine the mass and age of an isolated H~II region in the Virgo Cluster from its measured H$\alpha$ and $V$ band luminosities.  Thus far,  studies of these sorts of objects have focused on what the star clusters' integrated light reveals about their formation histories, and have been unable to resolve the clusters into individual stars or stellar components. With the discovery of more and more young, faint clusters in tidal tails of merging galaxies, intracluster space and the intergalactic medium, there arises a need for information about the nature of the forming stars. Yet, there is an unavoidable, inherent uncertainty in the properties of lower mass clusters derived from models that assume a fully populated IMF \citep{cervino}. Resolving the star clusters presents an opportunity to examine their populations in greater detail. 

We organize this paper as follows: Section 2 contains a description of our observations and measurements; Section 3 presents the results and model comparisons; Section 4.1 explores comparisons between these intergalactic H~II regions and other extragalactic and Galactic young stellar populations; in Sections 4.2 and 4.3, we discuss the gas properties in the vicinity of the intergalactic H~II regions and examine the likely fates of the stellar associations; and Section 5 briefly summarizes our results.  The Appendix contains the new calculation of the conversion factor between the ionizing luminosity and the total number of main sequence O stars contained within an H~II region following the methodology of Vacca (1994). 

\section{Observations and Measurements}

\subsection{HST Observations}

{\it{HST}} ACS/HRC observations were carried out in October 2004. The targets were observed in three filters: F250W (5808 s); F555W (2860 s); and F814W (2892 s), covering a wide wavelength range chosen to sample the shape of the ultraviolet to optical SED. Throughout this paper, we will refer to the F250W filter as UV, F555W as V, and F814W as I in the interest of brevity and simplicity. Each orbit was split into two exposures separated by a small dither, resulting in  $5\sigma$ limiting AB magnitudes of 28.0 in UV, 28.9 in V, and 29.0 in I.  The images were processed with the CALACS calibration pipeline in order to remove the instrumental signature, and then combined using the {\sl Apsis} pipeline \citep{bambm03} resulting in cleaned and combined single images in each filter at each pointing at the optimal pixel scale of ACS/HRC, $0\farcs025$ pixel$^{-1}$. We create a detection image by summing the images in the three filters. Due to the paucity of known point sources in our HRC images, we measured the PSF from HRC images  taken in the same cycle and in the same filters of the center of 47 Tuc, using at least 50 stars per image.  The point sources in each image have an average Gaussian FWHM of 2.6 pixels, independent of the 3 filters investigated.  At the adopted distance to NGC 1533, assumed to be 19.4 Mpc \citep{deGraaf07}, this FWHM corresponds to a physical size of 6.1 pc. 

Figure 1 shows the ACS/HRC detection images of the OB associations powering the isolated H~II regions along with their locations in the R-band continuum SINGG image of NGC 1533 \citep{SINGG};  we show the H~I distribution \citep{emma03} with contours.  The half-light radii of associations 1, 2, and 5 are 7.9, 9.0, and 6.2 pixels respectively, corresponding to physical sizes of 18.5, 21.1, and 14.5 pc.  At a distance of 19.4 Mpc, $1$ arcsecond corresponds to $94$ parsecs. Three to nine clumps, most close in size to the image PSF, comprise each association. While multiple sources bright in UV and V surround associations 1 and 2, no such objects appear within the same proximity to association 5.

\subsection{Source Detection and Definitions}

\begin{figure*}
\begin{center}
\plotone{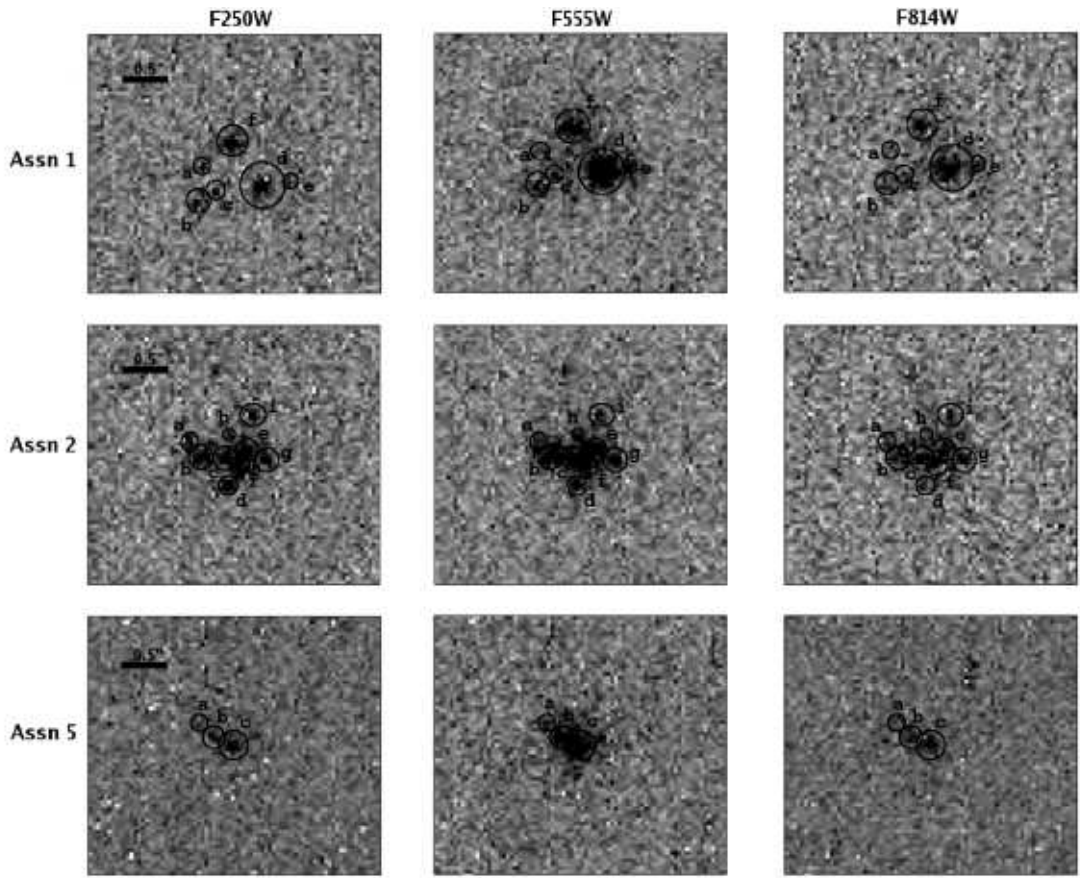}
\end{center}
\figcaption{Cut-out images in each HRC filter of the three isolated associations and their components (which we refer to as ``clumps"), circled and labeled. The image scale is shown in the 
top left corner. F250W is a UV bandpass, while F555W and F814W closely correspond to V and I bands, respectively. We present a detailed description of source detection and definition in Section 2.2. }

\end{figure*}

\begin{figure*}
\begin{center}
\plotone{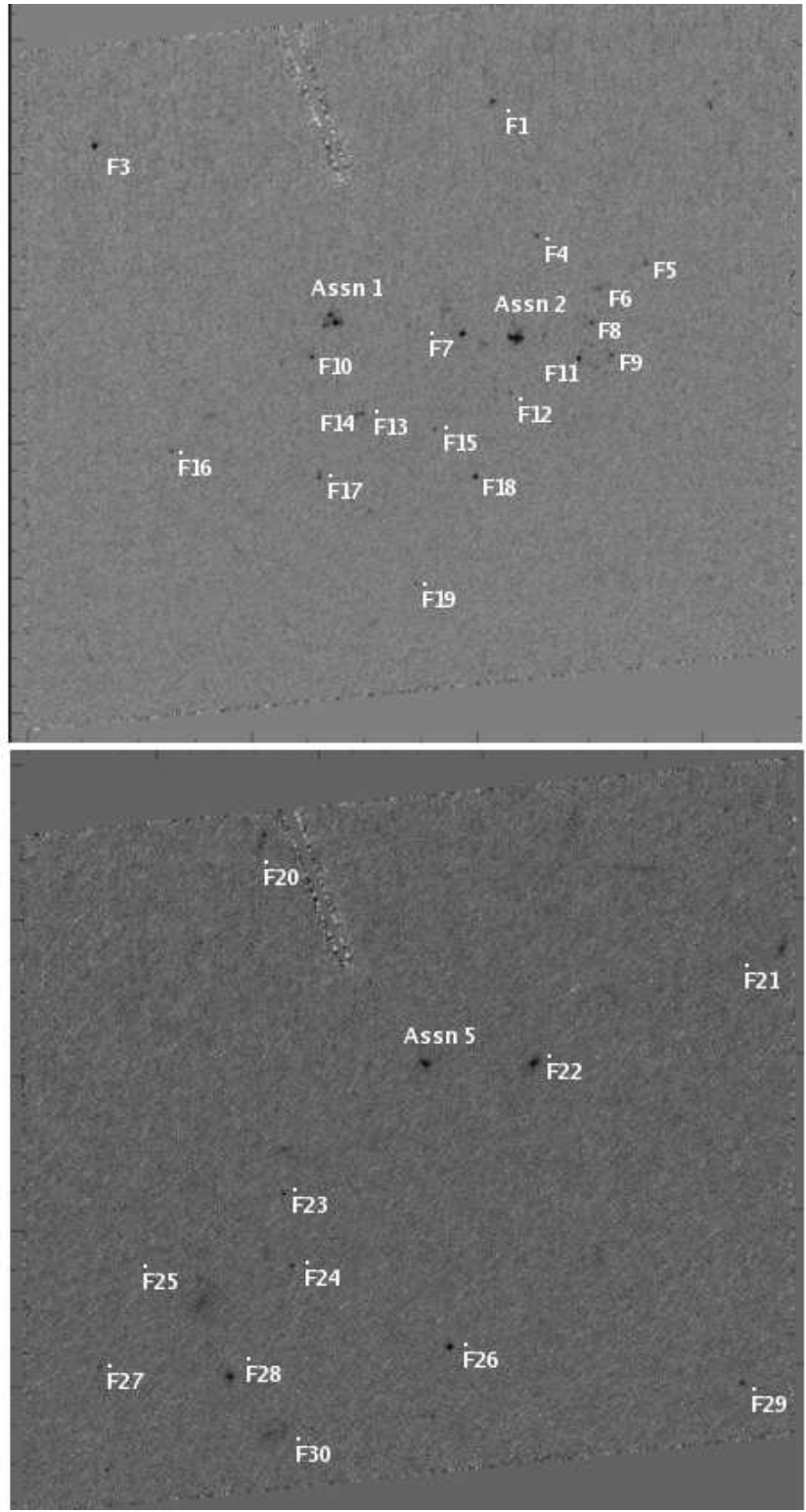}

\end{center}
\figcaption{HRC detection images show the full fields containing the isolated stellar associations 1,  2, and 5. The 30 detected field objects are labeled according to their names in Tables 2, 3 and 4. }

\end{figure*}

\begin{figure*}
\begin{center}
\plotone{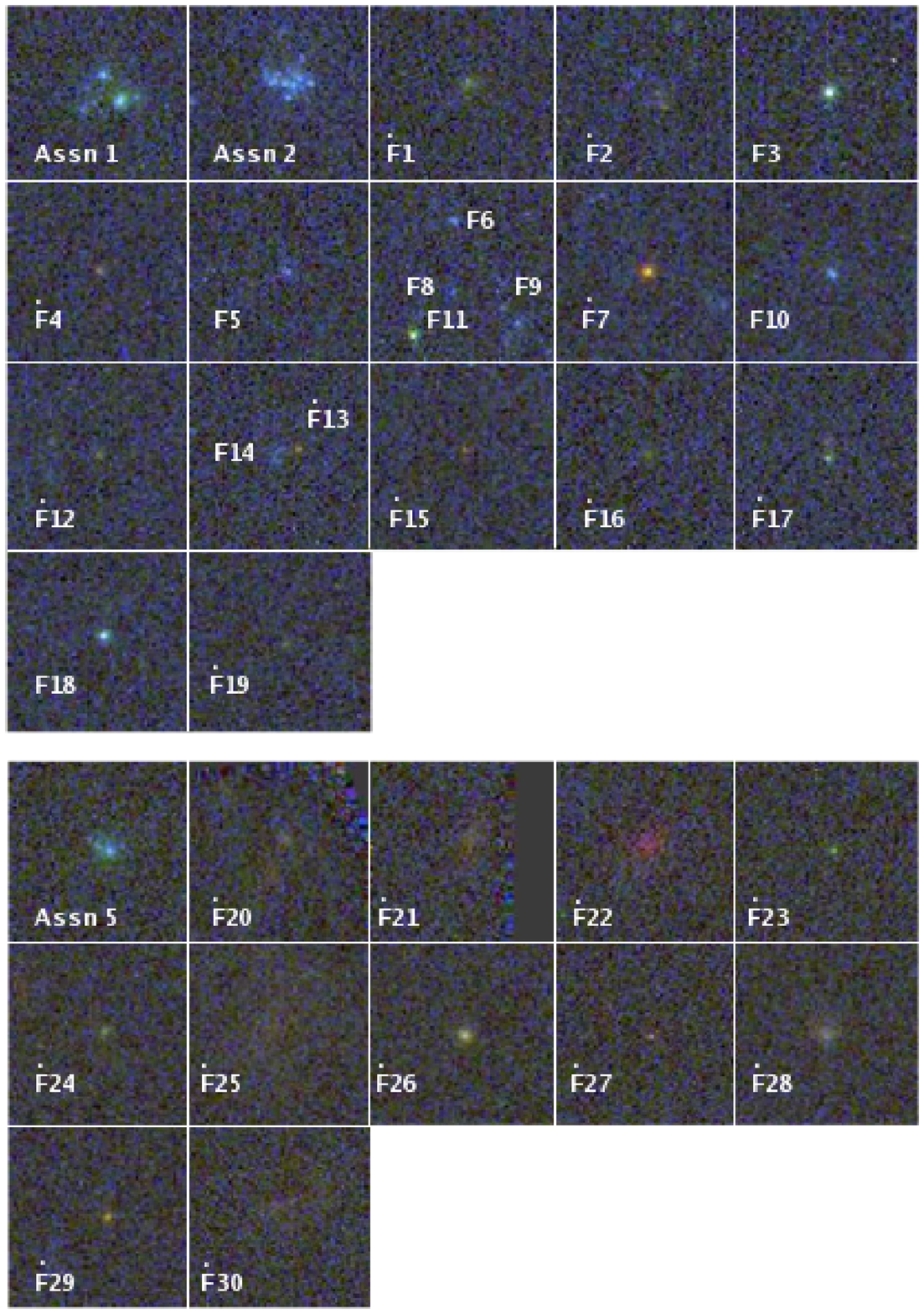}
\end{center}

\figcaption{Three-color images of associations and field objects where blue, green and red represent UV (F250W), V (F555W), and I (F814W), respectively. The objects are labeled with names corresponding to those given in Tables 2, 3, and 4. }
\end{figure*}

We catalogued the sources in our images by applying the SExtractor code \citep{bertin}, as implemented in the {\sl Apsis} pipeline, to the relevant detection images. We eliminated spurious sources (e.g. single pixel detections, image edge artifacts) and  objects detected in only one band by inspecting the SExtractor detections by-eye. Because the bulk of the following analysis requires color measurements, single-band detections (4 sources) are not of particular usefulness to us, though they may indeed be real. Of the initial 30 sources identified by SExtractor in the image containing associations 1 and 2, we include 19 in this analysis. Of the initial 36 sources identified by SExtractor in the image containing association 5, 11 appeared to be genuine detections, including one single-band detection (F814W) of an obvious background galaxy. Figure 1 shows the HRC detection images containing the labelled isolated associations. SExtractor split association 1 into three separate objects, and detected associations 2 and 5 as single objects. Further examination revealed each association to be composed of multiple clumps insufficiently deblended by SExtractor. Within the associations, we separated the components using IRAF contour plots with levels of 2$\sigma$, 4$\sigma$, 6$\sigma$ and 8$\sigma$, defining an individual component as having at least two contours, corresponding to at least 4$\sigma$ above the background. These criteria resulted in six components in association 1, nine components in association 2, and three components in association 5 (see Figure 2). Finally, using the IRAF task IMEXAM, we fitted Gaussian distributions to all field objects and clumps in order to determine their FWHMs. 

There are three main types of objects to which we will refer throughout this paper: the stellar associations powering the total H~II regions ($\equiv $associations), the components of the associations ($\equiv $clumps), and the field sources in the vicinity of the associations ($\equiv$ field objects). Furthermore, we assume that all of the extended and red field objects in the vicinity of the associations  are background galaxies (see Section 2.4 and Table 4), and thus exclude them from most of this analysis. We address background galaxy contamination of the rest of the field objects in Section 2.4. The images of the objects included in this study are presented in Figures 1 - 4.  Figure 1 gives an idea of the environment containing the intergalactic associations. All three lie in a low column density H~I ring, with projected radii 19 - 31 kpc from the nucleus of NGC 1533. Figure 2 displays close-ups of the clumps in the three bands, UV (left column), V (center column), and I (right column), circled and labeled. The letters in the images correspond to the names in Table 2.  Figure 3 shows the two HRC fields (in the three combined filters), with labels next to each object.   Finally, Figure 4 shows three-color image close-ups of the associations, clumps, and field objects in the two sets of HRC images. In these color close-ups, blue, green, and red represent the $UV$, $V$, and $I$ bands, respectively. 

\begin{deluxetable*}{cccccccr}
\tabletypesize{\small}
\tablewidth{0pc}
\tablecaption{\small Photometry Results and Sizes for Isolated Associations and Their Components}
\tablehead{\colhead{Assn}&\colhead{}&\colhead{F$_{\lambda}$ (F250W)}&\colhead{V-I\tablenotemark{a}}&\colhead{UV-V}&\colhead{m$_{V}$}&\colhead{M$_{V}$}&\colhead{FWHM}\\
\colhead{}&\colhead{}&\colhead{$10^{-18}$ cgs\tablenotemark{b}}&\colhead{Vega}&\colhead{AB}&\colhead{}&\colhead{}&\colhead{pc}}
\startdata
     1& Total & $8.13\pm0.78$& -0.12$^{+.10}_{-.10}$& -0.21$^{+ .11 }_{- .12 }$& 23.38$^{+.04}_{-.04}$&-8.06&61\\
\hline

&a&$0.24\pm0.04$& -0.69$^{+.45}_{-.52}$& -0.86$^{+ .25 }_{- .30}$ &27.85$^{+.17}_{-.20}$&-3.59&$3.39$\\
&b&$0.52\pm0.07$& -0.22$^{+.18}_{-.21}$&  -0.48$^{+ .16  }_{- .18}$&26.64$^{+.09}_{-.10}$&-4.80&$4.91$\\
&c&$0.18\pm0.04$& -0.33$^{+.27}_{-.35}$&  -0.18$^{+ .27  }_{- .34}$&27.51$^{+.13}_{-.15}$&-3.93&$5.48$\\
&d&$2.17\pm0.14$& -0.46$^{+.05}_{-.05}$& \thinspace\thinspace  0.33$^{+.07 }_{- .07}$ &24.27$^{+.02}_{-.02}$&-7.17&$9.40$\\
&e&$0.21\pm0.04$& -0.37$^{+.20}_{-.23}$&  \thinspace\thinspace0.07$^{+.22  }_{-.27}$ &27.06$^{+.09}_{-.10}$&-4.38&$4.00$\\
&f&$1.95\pm0.09$& -0.45$^{+.10}_{-.10}$&   -0.66$^{+ .06  }_{- .06}$&25.38$^{+.04}_{-.04}$&-6.06&$4.98$\\
&DLC&$2.63\pm0.78$& -0.25$^{+.19}_{-.22}$& \thinspace\thinspace0.33$^{+.29}_{-.39}$& 24.04$^{+.09}_{-.10}$&-7.40&--\\
\hline\hline
    2 & Total & $11.2\pm0.77$&  \thinspace\thinspace0.08$^{+.09}_{-.10}$&-0.67$^{+ .08 }_{- .09}$&23.51$^{+.04}_{-.05}$&-7.99&67\\
      \hline
       
&a&$0.26\pm0.04$& -0.02$^{+.22}_{-.27}$&  -0.54$^{+ .21 }_{- .24}$ &27.47$^{+.12}_{-.14}$&-3.97&$4.44$\\
&b&$1.18\pm0.07$& -0.31$^{+.09}_{-.10}$& -0.47$^{+ .07 }_{-.08 }$&25.73$^{+.04}_{-.04}$&-5.71&$9.66$\\
&c&$1.40\pm0.07$& -0.16$^{+.08}_{-.09}$&  -0.67$^{+ .06  }_{- .07}$&25.75$^{+.04}_{-.04}$&-5.69&$8.67$\\
&d&$0.62\pm0.05$& -0.86$^{+.23}_{-.28}$& -0.83$^{+ .11}_{- .12 }$&26.80$^{+.07}_{-.08}$&-4.64&$5.31$\\
&e&$1.25\pm0.07$& -0.25$^{+.09}_{-.10}$&  -0.62$^{+    .07  }_{- .08 }$&25.82$^{+.04}_{-.05}$&-5.62&$9.60$\\
&f&$0.93\pm0.05$& -0.45$^{+.11}_{-.12}$&-0.74$^{+ .08 }_{-.08}$&26.27$^{+.05}_{-.05}$&-5.17&$<4.70$\\
&g&$0.85\pm0.07$& -0.17$^{+.13}_{-.15}$&-0.64$^{+ .10 }_{- .11}$&26.26$^{+.06}_{-.07}$&-5.18&$4.83$\\
&h&$0.17\pm0.04$& -0.09$^{+.23}_{-.28}$&-0.11$^{+ .27 }_{- .34 }$&27.47$^{+.12}_{-.13}$&-3.97&$<4.70$\\
&i&$0.58\pm0.07$&   \thinspace\thinspace 0.19$^{+.20}_{-.23}$& -1.03$^{+ .17}_{-.18}$ &27.06$^{+.12}_{-.13}$&-4.38&$5.03$\\

   &DLC&$3.96\pm0.77$& \thinspace\thinspace0.17$^{+.17}_{-.19}$&-0.26$^{+.20}_{-.24}$  &24.19$^{+.10}_{-.11}$&-7.25&--\\
\hline\hline

     5 &Total & $3.51\pm0.66$& -0.22$^{+.15}_{-.17}$&  \thinspace\thinspace0.02$^{+ .20 }_{-.23}$ &24.06$^{+.06}_{-.06}$&-7.38&40\\
\hline

&a&$0.15\pm0.05$&  \thinspace\thinspace 0.46$^{+.29}_{-.36}$&  -0.55$^{+ .36  }_{-.48}$&28.05$^{+.20}_{-.25}$&-3.39&$4.46$\\
&b&$0.61\pm0.07$& -0.45$^{+.12}_{-.13}$&  \thinspace\thinspace0.02$^{+.13 }_{- .15}$&25.96$^{+.05}_{-.06}$&-5.48&$<4.70$\\
&c&$1.57\pm0.10$& -0.54$^{+.10}_{-.11}$& -0.33$^{+ .08 }_{-.08 }$&25.28$^{+.04}_{-.04}$&-6.16&$9.54$\\
&DLC&$1.33\pm0.66$&-0.54$^{+.25}_{-.32}$& \thinspace\thinspace0.74$^{+.44}_{-.75}$ &24.37$^{+.10}_{-.11}$&-7.07&--\\
\enddata
\tablenotetext{a}{\scriptsize V - I colors have been transformed to  Vega Magnitudes in the standard Johnson-Cousins filter system using the prescription of Sirianni et al. (2005). $V$ band magnitudes are technically in units of Vega magnitudes, although magnitude systems converge in the $V$ band, making the given values roughly correct for the AB magnitude system as well.  The UV - V colors are left in AB magnitudes, as there is no calibration for the conversion of the F250W filter to Vega magnitudes. Absolute magnitudes were calculated using a distance modulus of 31.44 (Blakeslee, private communication). For each association, V band magnitudes were corrected for nebular emission ([OIII]$\lambda\lambda$4959,5007).}
\tablenotetext{b}{\scriptsize ergs s$^{-1}$ cm$^{-2}$ \AA$^{-1}$}

\end{deluxetable*}

\subsection{Aperture Photometry}

We performed circular aperture photometry on each source using the IDL routine APER. Aperture sizes for each association and field object were determined by examining where the curve of growth (in flux vs. aperture radius) began to flatten, whereas aperture sizes for the clumps were determined by eye.  For each association and field object, sky annuli were positioned approximately $1$ arcsecond away from each of the objects' centers. Because the clumps are so closely spaced in the image, we subtracted the sky measurement outside the corresponding association as a best estimate.  As evidenced by Figure 2, the circular apertures containing the clumps do not account for all the light in the association. We chose the aperture sizes to contain the maximum amount of emission while not overlapping with other apertures.  By masking out the apertures containing the clumps, and running APER on what remained, we determined fluxes for the diffuse light components (DLC) of each association. This component simply represents the light losses resultant from the small apertures, and thus may not be genuine diffuse emission. Thus, the DLC is essentially the difference between the total association flux, and the sum of the fluxes of the individual clumps.

A correction for foreground extinction, based on the \cite{Schlegel98} extinction maps, (foreground E(B-V) = 0.016), was applied according to the parameterization of \cite{ccm}, including the update for the near-UV given by \cite{odon}. Given the location of the isolated associations far outside the optically luminous area of NGC 1533, we excluded any correction for the internal extinction. Additionally, fluxes for the associations were corrected for the strong emission lines [{\ion{O}{3}}]$\lambda\lambda 4959,5007$ present in the F555W bandpass. We used the STSDAS package SYNPHOT to simulate photometry using an actual spectrum of association 1 (to be presented in a future paper) both with and without the nebular emission lines. The procedure for this correction is as follows: (1) We measured the H$\alpha$ flux for all three H~II regions in the NGC1533 SINGG narrow-band, continuum-subtracted image; (2) We measured the H$\alpha$ emission line flux in the spectrum of association 1; (3) We divide quantity (1) by quantity (2)  for each association to obtain a flux correction factor for each individual association; (4) Using SYNPHOT, we measure the total flux  of association 1's spectrum contained in the F555W bandpass; (5) We subtract the emission lines in this area of spectrum, and remeasure the total flux  in the same area with SYNPHOT; (6) We subtract (5) from (4) to get an uncalibrated flux of just the emission lines in the F555W bandpass; (7) Finally, we calibrate and scale the total nebular emission-line flux from (6) by the individual correction factor determined in (3). 

These emission-line corrections were 0.26, 0.17, and 0.27 AB magnitudes ($17\% - 28\%$ of the total flux) for associations 1, 2 and 5, respectively. We note that these emission line corrections assume that the flux ratios of H$\alpha$ to [O~III]  emission lines are the same for each association. The error due to potential reasonable variations in these line ratios is negligible when compared to the total corrections. Furthermore, this correction has the effect that the associations are redder in V-I after correction than one would expect based on the V-I colors of the individual clumps, which were not corrected for nebular emission due to the limited ($\sim$1.5 arcsecond) resolution of the SINGG H$\alpha$ images. 

Finally, for comparison with population synthesis models and previous work, we transformed the ST magnitudes to the Vega magnitude system, and the filters to the standard Johnson-Cousins system following the method outlined by \cite{sirianni} for both the F555W and F814W data. Because magnitude systems converge in the $V$ band, and because the F555W filter is quite similar to the Johnson-Cousins $V$ band, its transformation produces virtually no change. Furthermore, the F250W filter has no Johnson-Cousins counterpart, and it is impractical to convert AB magnitudes to Vega magnitudes in this part of the spectrum. Throughout this paper, we will quote V-I colors  in terms of Vega magnitudes in the standard Johnson-Cousins filters. This magnitude system allows for the greatest ease in comparing our results with both previous work and existing population synthesis models. However, UV-V and UV-I colors and UV magnitudes will remain in the HRC filter bandpasses and in AB magnitudes.  The faintness of the clumps and field objects combined with considerable pixel-pixel noise in the read noise limit produces errors in the photometry as high as 0.25 magnitudes.  The resulting Vega magnitudes are presented in Tables 2, 3, and 4 as well as the V-I  and UV-V colors,  along with the measured FWHM of each source.

\subsection{Source Properties and Contamination by Background Galaxies}

Our sample of field objects and clumps totals 48 individual sources altogether,  34 of which are located in the 3 kpc $\times$ 3 kpc field around associations 1 and 2.    We estimate that the associations span 40 - 75 parsecs in their full visible extent, similar to large Milky Way stellar associations. Figure 3 gives the location of all objects identified in the two HRC images, with the field containing associations 1 and 2 on the top frame, and that of association 5 on the bottom frame. Of the 48 individual sources, only 17 are resolved. 9 of the 11 field objects in the association 5 field are well-resolved extended sources. The V-I colors of the clumps range from -0.7 to 0.2 in V-I, and those of the field sources range from -0.9 to 1.8. If all the sources were at the same distance as NGC 1533, the absolute $V$-band magnitudes range from -3.4 to -7.3.  

Almost certainly, some of the field objects in the HRC images represent distant background galaxies. From Hubble Deep Field galaxy number densities, we expect 1-2 $\times$ 10$^{5}$ galaxies per square degree in WFPC2, corresponding to a limiting I-band magnitude of 25-26 \citep{casertano00}. Field galaxy number densities are lower in the far-UV, roughly 3 $\times$ 10$^{4}$ galaxies per square degree to a limiting AB magnitude of 26.5 \citep{teplitz06}. Considering these number densities, we would then expect 5-10 red background galaxies to appear in our 29$\arcsec\times$26$\arcsec$ field of view, and 0-1 galaxies with emission in the UV. We detect and analyze 11 objects in the vicinity of association 5, none of which appear in the UV images, and the majority of which appear quite red in V-I. Of the 11 objects in the vicinity of association 5, six are very obviously extended or elliptical red sources, with V-I colors (Vega magnitudes) between 0.7 and 1.8 (as previously mentioned, one object is detected only in F814W). Three of the remaining five compact sources have V-I colors in a similarly red range, while two exhibit fairly blue V-I colors, 0.04 and 0.19. The lack of UV emission from these sources combined with the number being consistent with our prediction for background galaxy detections leads us to classify all 11 as background emitters. 

We detect and analyze 19 objects in the vicinity of associations 1 and 2. In contrast to the association 5 field objects, many of these field objects appear bright in UV and V, suggesting that the star formation in the SE part of NGC 1533's H~I ring extends beyond the initially detected isolated H~II regions. Furthermore, most of the surrounding blue field objects lie clustered within 500 parsecs of the nearest association (1 or 2), while the likely background galaxies are spread more evenly throughout the image. While we cannot say for certain which of the field objects around associations 1 and 2 are background galaxies, we can predict that the 10 sources not detected in F250W, 5 of which have V-I colors greater than 1.0, are more likely  to be distant background galaxies. These sources also appear to be more extended. Moreover, this number of sources is  consistent with the prediction based on the background galaxy density of deep WFPC2 images. 

We do note, however, that we are unable to distinguish extended likely background galaxies from potentially old ($\sim1$ Gyr), extended open clusters, which can have V-I colors up to 1.4 and absolute magnitudes ranging from -1 to -5.1 in the $V$ band \citep{lata02}. Nor can we differentiate unresolved likely background galaxies from globular clusters, which have V-I colors ranging from 0.85 to 2.65, and absolute $V$-band magnitudes from -1.5 to -9.5 \citep{harris96}. Still, such old, extended open clusters are rare in the Milky Way (e.g. \nocite{ladalada,lata02} Lata et al. 2002; Lada \& Lada 2003), and globular clusters tend to lie closer to the galaxy center (within $\sim20$ kpc) than the star formation seen here \citep{harris96}, although a few are as far as 100 kpc from Galactic center. Still another possibility remains that some of these objects could represent isolated red supergiants. 

Tables 3 and  4 present the photometric properties and optical sizes for all detected field objects. In these tables, we have identified those field objects which are the best candidates for distant background galaxies (extended, red, and/or no UV-emission) as $\dot{\rm F}$,  and those blue field objects which are most likely to be associated with the ongoing star-formation around NGC 1533 as F.  Figure 3 shows three-color (RGB) images of the field objects and background galaxies. The top panel shows the field objects surrounding associations 1 and 2, and the bottom panel displays the field objects of association 5. The labels on each image correspond to the names listed in Tables 3 and 4 (``Assn" for the association; ``F" for field object, and ``$\dot{\rm F}$" for likely background galaxy). We do include all of these objects in the subsequent analysis for completeness' sake (except where their red colors push them outside of adopted plotting ranges), and we urge the reader to consider the 10 ``dotted" objects in Table 3 and the 11 objects in Table 4 (also identified in the figures) as potential background galaxies. 

\begin{deluxetable*}{rcccccclr}
\tabletypesize{\small}
\tablewidth{0pc}
\tablecaption{\small Photometry Results and Sizes for Association 1 \& 2 Field Objects}

\tablehead{\colhead{ID}&\colhead{R.A.}&\colhead{Dec.}&\colhead{F$_{\lambda}$ (F250W)}&\colhead{V-I\tablenotemark{a}}&\colhead{UV-V}&\colhead{m$_{V}$}&\colhead{M$_{V}$}&\colhead{FWHM}\\
\colhead{}&\colhead{h m s}&\colhead{d m s}&\colhead{$10^{-18}$ cgs\tablenotemark{b}}&\colhead{Vega}&\colhead{AB}&\colhead{}&\colhead{ }&\colhead{pc}}
\startdata
$\dot{\rm F}$1& 04 10 14.54  &-56 11 44.84     &--&                 \thinspace\thinspace0.05$^{+.23}_{-.26}$&  --&25.52$^{+.14}_{-.16}$&-5.92&  11.38\\
 $\dot{\rm F}$2& 04 10 15.51  &-56 11 43.35     &--&                  \thinspace\thinspace1.38$^{+.32 }_{-.43}$&  --&26.63$^{+.28}_{-.39}$&-4.81&  7.81\\
 F3& 04 10 12.73  &-56 11 45.58     &$1.36\pm0.14$&        -0.23$^{+.04}_{-.04}$&\thinspace\thinspace1.17$^{+ .19    }_{- .23}$&23.92$^{+.03}_{-.03}$&-7.52&   5.91\\
 $\dot{\rm F}$4& 04 10 14.64  &-56 11 39.58     &--&       \thinspace\thinspace 1.01$^{+.17}_{-.19}$& --&26.55$^{+.14}_{-.16}$&-4.89&   $7.88$\\
 F5& 04 10 15.10  &-56 11 37.89     &$1.02\pm0.13$&       -0.40$^{+.34}_{-.45}$& -0.95$^{+ .20  }_{- .23 }$&26.35$^{+.15}_{-.17}$&-5.09&   $5.60$\\
 F6& 04 10 14.82  &-56 11 36.00     &$0.90\pm0.14$&       -0.61$^{+.36}_{-.51}$& -0.74$^{+ .22   }_{- .25}$ &26.27$^{+.15}_{-.17}$&-5.17&   $5.95$\\
 $\dot{\rm F}$7& 04 10 14.24  &-56 11 36.35     &--&                   \thinspace\thinspace2.03$^{+.10 }_{-.11}$& --&25.24$^{+.10}_{-.11}$&-6.20&   $6.02$\\
 F8& 04 10 14.80  &-56 11 35.18     &$0.56\pm0.11$&         --&        -1.07$^{+  .25}_{-.30}$       &27.12$^{+.17}_{-.20}$&-4.32&   3.97\\
 F9& 04 10 14.89  &-56 11 34.69     &$0.75\pm0.13$&        \thinspace\thinspace0.34$^{+.21}_{-.25}$& -0.67$^{+ .23  }_{-.26}$ &26.41$^{+.15}_{-.17}$&-5.03&   5.62\\
 F10& 04 10 13.55  &-56 11 36.42     &$1.59\pm0.14$&      -0.27$^{+.18}_{-.21}$& -0.86$^{+ .13 }_{- .14 }$&25.78$^{+.09}_{-.10}$&-5.66&   $7.15$\\
 F11& 04 10 14.74  &-56 11 34.77     &$0.35\pm0.14$&      -0.10$^{+.07}_{-.08}$& \thinspace\thinspace2.04$^{+.36  }_{-  .55 }$&24.56$^{+.04}_{-.04}$&-6.88&   $5.48$\\
 $\dot{\rm F}$12& 04 10 14.41  &-56 11 33.89     &--&       \thinspace\thinspace 0.79$^{+.24}_{-.30}$&--&26.95$^{+.20}_{-.25}$&-4.49&   $5.34$\\
 $\dot{\rm F}$13& 04 10 13.74  &-56 11 34.02     &--&                  \thinspace\thinspace 1.49$^{+.27}_{-.34}$&  -- &26.94$^{+.25}_{-.32}$&-4.50&   $5.64$\\
 F14& 04 10 13.70  &-56 11 33.91     &$0.68\pm0.17$&      -0.89$^{+.52}_{-.93}$&   -0.44$^{+ .31  }_{-  .39}$&26.28$^{+.19}_{-.23}$&-5.16&   $4.66$\\
 $\dot{\rm F}$15& 04 10 14.05  &-56 11 32.99     &--&                    \thinspace\thinspace1.36$^{+.27}_{-.34}$& -- &27.28$^{+.25}_{-.32}$&-4.16&   $5.60$\\
 $\dot{\rm F}$16& 04 10 12.86  &-56 11 33.76     &--&                    \thinspace\thinspace0.67$^{+.29}_{-.36}$& -- &26.61$^{+.24}_{-.30}$&-4.83&   $8.02$\\
 $\dot{\rm F}$17& 04 10 13.49  &-56 11 31.89     &--&     -0.32$^{+.20}_{-.24}$& --&25.77$^{+.09}_{-.10}$&-5.67&   $5.19$\\
 F18& 04 10 14.19  &-56 11 30.99     &$2.41\pm0.16$&    -0.27$^{+.08}_{-.09}$&-0.19$^{+ .08 }_{-.09  }$&24.66$^{+.04}_{-.04}$&-6.78&   $5.62$\\
 $\dot{\rm F}$19& 04 10 13.85  &-56 11 27.36     &--&    -0.01$^{+.27}_{-.33}$& --&26.92$^{+.16}_{-.19}$&-4.52&   $5.78$\\
\enddata
\tablenotetext{a}{\scriptsize V-I colors have been transformed Vega Magnitudes  in the standard Johnson-Cousins filter system using the prescription of Sirianni et al. (2005). Absolute magnitudes and physical sizes were calculated using a distance modulus of 31.44 (D = 19.4 Mpc, Blakeslee, private communication). }
\tablenotetext{b}{\scriptsize ergs s$^{-1}$ cm$^{-2}$ \AA$^{-1}$}
\end{deluxetable*}

\begin{deluxetable*}{lccccr}
\tabletypesize{\small}
\tablewidth{0pc}
\tablecaption{\small Photometry Results and Sizes for Association 5 Field Objects}
\tablehead{\colhead{ID}&\colhead{R.A.}&\colhead{Dec.}&\colhead{V-I}&\colhead{m$_{v}$}&\colhead{FWHM\tablenotemark{a}}\\
\colhead{}&\colhead{h m s}&\colhead{d m s}&\colhead{Vega}&\colhead{}&\colhead{pc}}
\startdata
$\dot{\rm F}$20& 04 10 16.52  &-56 06 24.07     &1.22$^{+.28}_{-.36}$& 25.27$^{+.25}_{-.33}$&  14.68\\
$\dot{\rm F}$21& 04 10 16.86 & -56 06 04.30&1.71$^{+.22}_{-.28}$& 25.64$^{+.21}_{-.27}$&  15.47\\
$\dot{\rm F}$22\tablenotemark{b}& 04 10 16.00 & -56 06 11.62&--& --& 17.48\\
$\dot{\rm F}$23& 04 10 15.06 & -56 06 18.80&0.04$^{+.16}_{-.18}$& 26.00$^{+.09}_{-.10}$&  6.95\\
$\dot{\rm F}$24& 04 10 14.77 & -56 06 17.57&0.19$^{+.16}_{-.18}$& 25.99$^{+.11}_{-.12}$&  9.99\\
$\dot{\rm F}$25& 04 10 14.48 & -56 06 20.22&1.56$^{+.21}_{-.25}$& 24.30$^{+.19}_{-.23}$&  24.73\\
$\dot{\rm F}$26& 04 10 14.66 & -56 06 10.95&0.85$^{+.07}_{-.09}$& 24.80$^{+.06}_{-.07}$&  9.91\\
$\dot{\rm F}$27& 04 10 14.04 & -56 06 23.07&2.25$^{+.28}_{-.38}$& 27.14$^{+.27}_{-.37}$&  5.89\\
$\dot{\rm F}$28& 04 10 14.20 & -56 06 18.31&0.71$^{+.11}_{-.11}$& 24.81$^{+.08}_{-.08}$&  17.90\\
$\dot{\rm F}$29& 04 10 14.02 & -56 06 16.17&1.34$^{+.13}_{-.13}$& 26.09$^{+.11}_{-.12}$&  6.36\\
$\dot{\rm F}$30& 04 10 14.96 & -56 06 00.15&1.79$^{+.32}_{-.43}$& 24.83$^{+.30}_{-.42}$&  24.97\\
\enddata
\tablenotetext{a}{\scriptsize Physical sizes are given in parsecs for straightforward comparison with clumps and other field objects. }
\tablenotetext{b}{\scriptsize We do not detect $\dot{\rm F}$22 in any filter other than F814W, although we consider it a real source, with an $I$ band magnitude of 23.14 (Vega magnitudes). }
\end{deluxetable*}

In Figures 5 and 6, we compare the individual sources' absolute magnitudes and V-I and UV-V colors with those of single star MK spectral types and a 50 M$_{\odot}$ instantaneous-burst Starburst99 population (see 3.1). In order to plot MK spectral types in M$_{V}$ versus UV-V, we converted their U-band absolute Vega magnitudes to the F250W filter in AB magnitudes using the SYNPHOT.  To arrive at Starburst99 UV-V colors we calculated the F250W AB magnitude using the output synthetic spectrum along with SYNPHOT.  In many cases, the colors and absolute magnitudes of the objects are well-matched by single O or B stars. This similarity does not prove that field objects and clumps are single stars, but that O and B stars dominate the radiation output in many cases. In other cases, the expected magnitude-color evolution of a Starburst99 population (run for an instantaneous burst Salpeter IMF with$\alpha = 2.35$, $M_{up} = 100$, $M_{low} = 0.5$,  Padova group isochrones, and  Z$=0.4$Z$_{\odot}$) with an initial mass of 50 M$_{\odot}$ matches the data. We note that  a 50 M$_{\odot}$ population with a 100 M$_{\odot}$ upper mass limit is unphysical, and results in fractional stars. We have simply used an initial starburst mass of  50 M$_{\odot}$  to appropriately scale the absolute magnitude and illustrate a degeneracy. The colors are unaffected by the mass of the initial starburst. We discuss the implications of fractional stars and stochastic IMFs in section \ref{stoceffs}. For completeness, we have included the bluest (in V-I) background galaxy candidates in Figure 5, using NGC 1533's distance modulus to calculate their absolute magnitudes.  The reddest candidates do not lie within the ascribed color range. 

\begin{figure}[t]
\begin{center}
\plotone{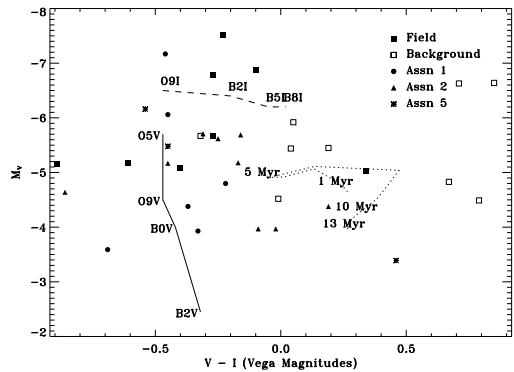}

\end{center}
\figcaption{Color-magnitude diagram for the blue field objects and clumps, with V-I in Vega magnitudes. The dashed and solid lines shown are calibrations of MK spectral types from Schmidt-Kaler (1982). The dotted line represents the V-I color evolution of a Starburst99 50 M$_{\odot}$ population with Z$=0.4$Z$_{\odot}$. The population synthesis model was run for an instantaneous burst Salpeter IMF ($\alpha = 2.35$, $M_{up} = 100$, $M_{low} = 0.5$), and employs Padova group isochrones.}

\end{figure}

\begin{figure}[t]
\begin{center}
\plotone{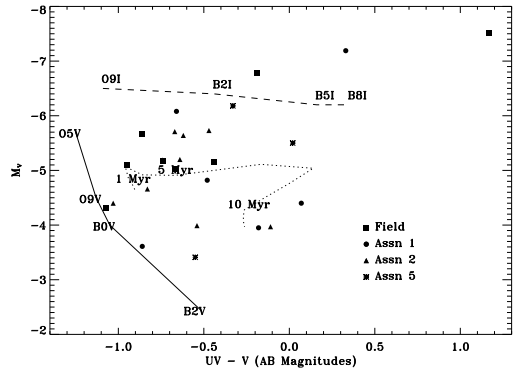}

\end{center}

\figcaption{Color-magnitude diagram for the field objects and clumps, with UV-V in AB magnitudes.  The dashed and solid lines shown are calibrations of MK spectral types from Schmidt-Kaler (1982). The dotted line represents the UV-V color evolution of the same Starburst99 population shown in Figure 5. }
\end{figure}

\section{Stellar Populations of Associations}

Ideally, stellar population synthesis models, along with integrated colors and luminosities, can provide a solution for the masses and ages of the stars populating the clusters. These models aim to account for most of the known stellar evolutionary processes, and integrate over various sets of assumed initial conditions. In the following analysis, we describe two approaches we took in an effort to constrain the stellar populations of the isolated associations. We compare the associations' colors and luminosities to output from Starburst99 models \citep{starburst99}, and we perform a simple, independent calculation using the associations'  total UV and H$\alpha$ luminosities. Both approaches employ single-age, single-metallicity models with a Salpeter IMF. While the Starburst99 model accounts for details of stellar evolution, it cannot account for anything but a scaled-down IMF over the full range of stellar initial masses. Our independent calculation is more quantized with regard to the stellar populations, but does not account for some of the details of stellar evolution. In section \ref{stoceffs} we discuss the limited applicability of both models to our data given the low luminosities of the isolated associations, and masses below what is required for a fully-populated IMF.  

\subsection{Comparison with Starburst99 Model}

To accurately designate a metallicity for the Starburst99 model, we used an optical spectrum of association 1, estimating its metallicity to be roughly $\sim 0.4 Z_{\odot}$ (Werk et al., in preparation).  Figure 7 plots various diagnostics from ages 1 - 16 Myr for a Starburst99 synthetic population for Z$=0.4$Z$_{\odot}$, an instantaneous burst of 10$^{6}$M$_{\odot}$, and a Salpeter IMF with a lower limit of $0.5$ M$_{\odot}$ and an upper limit of $100$  M$_{\odot}$.  It also shows the  corresponding values calculated for associations 1, 2 and 5. The error bars indicate photometric errors.

In Figure 7a, we show the temporal evolution of the ratio Log $(F_{H\alpha}/F_{UV})$ for the Starburst99 model and the three associations.  H$\alpha$ and UV luminosities serve as reasonable star formation indicators \citep{kennicutt98}. While H$\alpha$ traces the ionizing radiation from stars more massive than $15M_{\odot}$, the UV luminosity originates from stars more massive than $3M_{\odot}$. The ratio Log $(F_{H\alpha}/F_{UV})$ decreases steeply with time, as massive ionizing stars die out, and late-type O stars and B stars produce the majority of the UV flux. The best-match ages  to the evolution of  Log $(F_{H\alpha}/F_{UV})$ are 3.2 Myrs, 4.5 Myrs, and 3.0 Myrs for associations 1, 2, and 5 respectively.  In this plot, the horizontal error bars represent the minimum possible error given the photometric errors. 

\begin{figure*}
\begin{center}
\plotone{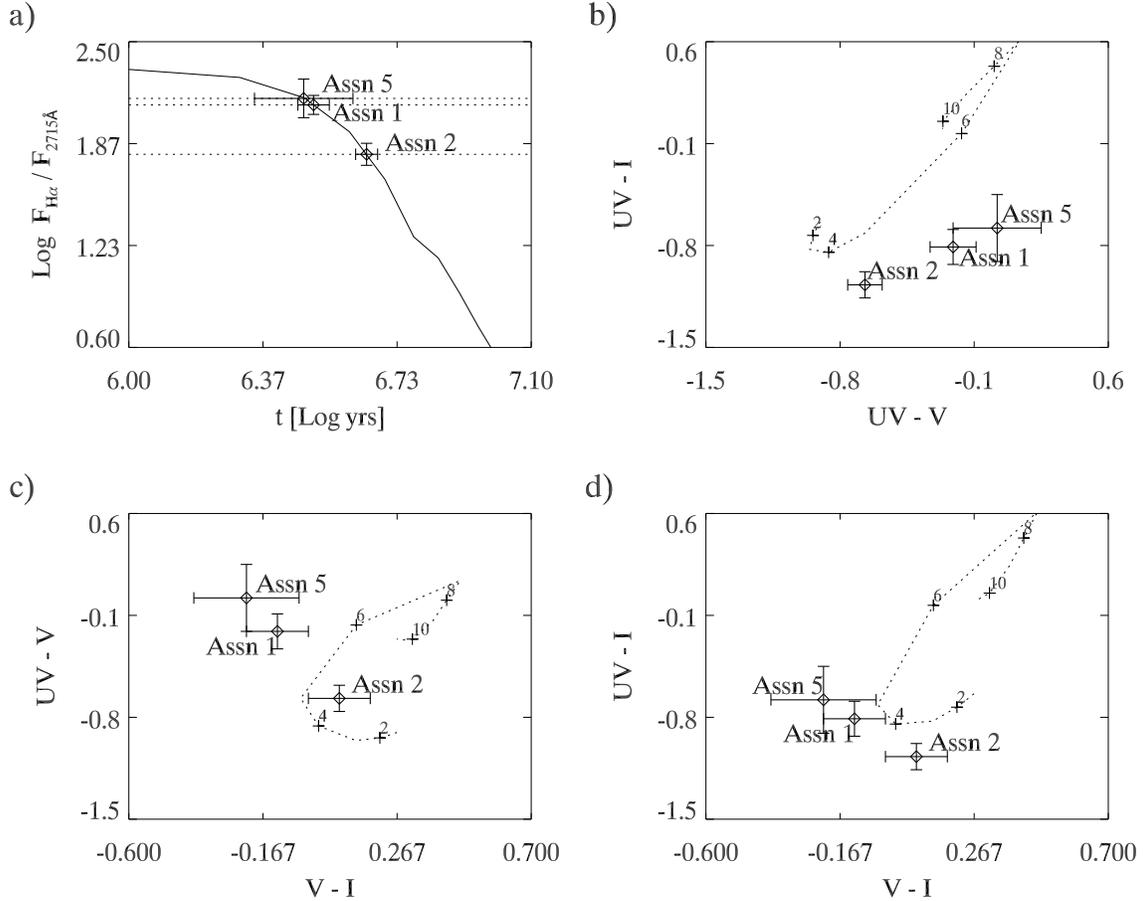}

\end{center}
\figcaption{Evolution with time of the H$\alpha$ to UV ratio and relevant optical color-color diagrams for the Starburst99 stellar population synthesis model at Z$=0.4$Z$_{\odot}$. The model was run for an instantaneous burst Salpeter IMF ($\alpha = 2.35$, $M_{up} = 100$), and employs Padova group isochrones.  a) The H$\alpha$ to UV ratio of each association with a symbol at the intersection point of the Starburst99 model.  The inferred ages are 3.2 Myr, 4.5 Myr, and 3.0 Myr for associations 1, 2, and 5 respectively. The errors in the best-match ages correspond to the uncertainty due to the photometric errors alone, and do not include stochastic effects or model errors. b) UV-I versus UV-V; the dotted line shows Starburst99 values for ages between 2 - 10 Myrs  c) UV-V versus V-I  d) UV-I versus V-I}

\end{figure*}

The evolution of  UV-V, UV-I, and V-I colors of the same Starburst99 population is plotted in three color-color diagrams for the same range in ages in Figures 7b, 7c, and 7d.   The Starburst99 UV magnitudes in the HRC F250W bandpass were generated by the SYNPHOT package for the output Starburst99 spectra. Seen in 7b, 7c, and 7d, associations 1, 2, and 5 are consistently bluer than the model colors, except for association 2 in V-I. Thus, we cannot constrain the ages using simply the colors of the associations.  As we will discuss, the inability of the Starburst99 model to match the association colors most likely arises from small number sampling in which a few early-type stars dominate the output radiation. An incompletely-sampled IMF results in a significant scatter of colors at a given age \citep{chiosi88}, and would explain the discrepancies seen in figures 7b, c, and d. 
 
 From the Starburst99 model, we can also derive mass estimates based on the absolute $V$ band magnitudes  and ages of the associations. At a particular age, a 10$^{6}$ M$_{\odot}$ Starburst99 stellar population has a predicted absolute $V$ band magnitude, M$_{V,S99}$. We then use a simple scaling relation between mass and absolute magnitude to derive the masses of the associations: M = 10$^{6 + 0.4(M_{V,S99} - M_{V})}$, where M is the mass of the stellar population and M$_{V}$ is the absolute magnitude of the association given in Table 2. The total stellar mass of each association is 1490, 1850, and 840 M$_{\odot}$ for associations 1, 2 and 5 respectively. We repeat this calculation for the I-band magnitudes. The total stellar mass from the I-band magnitude of each associations is 1240, 2030, and 610 M$_{\odot}$ for associations 1, 2 and 5, very similar to the results of the V-band calculation. We note that this particular way of estimating the mass assumes a scaled-down IMF containing the full-range of stellar masses from 0.5 M$_{\odot}$  to 100 M$_{\odot}$. Indeed, if we were to try to quantize the numbers of stars by their spectral type we would find only fractional stars for the highest masses. Table 5 includes columns for the ages and masses derived using the Starburst99 model.

 Although these plots  and derived parameters are instructive, they are limited by the low luminosities of the clusters.  A very basic limit for the application of evolutionary synthesis data to observational data, termed the ``lowest luminosity limit'' (LLL) by \cite{cervino}, is that {\it{the total luminosity of the cluster modeled must be larger than the individual contribution of any of the stars included in the model}}. Using Figure 2 from Cervino \& Luridiana (2004) for Z$=0.4$Z$_{\odot}$, we compare the absolute magnitudes of the associations with the LLL in both the V and I bands. For young star clusters ($\sim5$ Myrs), the minimum V and I band luminosities for which population synthesis models are valid are -10.0 and -10.5, respectively. In the UV, the LLL would be fainter, and the effects of undersampling will be less. Since the associations have absolute V and I band magnitudes significantly fainter than the limits, it is unlikely we are completely sampling the stellar population. Stochastic effects then will impact the data interpretation, and we explore these effects more thoroughly in Section \ref{stoceffs}.

\subsection{Independent Calculation of Cluster Ages and Masses}

We further constrain the stellar populations of the associations by performing an independent estimate of the total number of ionizing stars present in each association. We extrapolate this calculation to determine total masses and ages of the associations. We begin by converting an H$\alpha$ luminosity to an emission of ionizing photons,  Q$_{0} = 7.33\times10^{11}$ L$_{H\alpha}$ s$^{-1}$ with L$_{H\alpha}$ in ergs s$^{-1}$ cm$^{-2}$ \AA$^{-1}$  \citep{osterbrock}. Models then predict the number of ionizing photons released per unit time from a certain spectral type (ST) of a massive star, usually O7V.       By dividing the two quantities, one arrives at the number of  ``equivalent" stars of the given subtype. This number of  ``equivalent" ionizing stars, while a simple characterization of the stellar population, does not represent the {\it{total}} number of ionizing stars. To determine the total number of ionizing stars, one must integrate over an initial mass function, and use models that predict how the output of ionizing photons varies with stellar mass. In this section, we give the results of our calculation. In the appendix, we include a detailed, step-by-step description of the calculation, including all relevant equations and parameters.  Essentially, our method follows the calculations outlined in \cite{vacca94}, but uses the recently updated T$_{eff}$ - spectral type calibration of \cite{martins}.

For associations 1, 2, and 5 we determine the upper limits of the present-day mass function to be 35 M$_{\odot}$, 25 M$_{\odot}$, and 40 M$_{\odot}$, respectively.  The method we use, as described in the appendix, compares the total UV and H$\alpha$ fluxes of the associations and the relative contributions from O and B stars. Essentially, we are modeling the UV flux as the sum of the flux from both the O and B stars. We ultimately determine what value of stellar mass corresponds to a present-day mass function upper limit that can account for both the UV and H$\alpha$ fluxes that are observed. We use these upper masses to calculate a limit on the lifetimes of the clusters. For initial stellar rotation velocities between 0 and 300 km s$^{-1}$, a 35 M$_{\odot}$ star burns hydrogen for 4.7 - 5.9 Myr, a 25 M$_{\odot}$ star burns hydrogen for 5.9 - 7.4 Myr, and a 40 M$_{\odot}$ star burns hydrogen for 4.2 - 5.1 Myr  \citep{mandm}.  We obtain the total masses of the associations by integrating the present-day mass function, and using the upper mass cut-offs given above and lower mass limit of 0.5 M$_{\odot}$. The masses of associations 1, 2, and 5 calculated by this method are 2700, 7200, and 970 M$_{\odot}$, respectively. We note that this method of calculating the masses does not assume a scaled-down full IMF, as the Starburst99 mass estimate does, but rather uses a present-day mass function. We integrate only to the present-day upper mass limit, resulting in a more physical calculation of mass, minimizing the number of fractional stars. The  number of ``equivalent" O7V stars, total number of ionizing O stars, the total mass, and the derived ages from both the Starburst99 model and this independent calculation are given in Table 5 for each isolated H~II region. To abbreviate the names of the two methods, we refer to the Starburst99 derived parameters with S99, and the calculation using the method and models of \cite{vacca94,vacca96,martins} as VM05. The range in ages given in Table 5 for VM05 correspond to stellar lifetimes without rotation, and the minimum error contributed by the uncertainty in the photometry. 

There is general qualitative agreement between the corresponding quantities, yet some differences are clear. The ages determined with the VM05 calculation are systematically larger than those determined from Starburst99.  Neither model accounts for stochastic effects upon the stellar masses present, so the differences in ages and masses most likely reflect the difficulty of modelling the integrated properties of an undersampled stellar population. Association 2 is the most discrepant across the separate calculations, particularly in mass. 

A comparison between the number of  equivalent O7V stars and the total O star population is illustrative of the limited information gained from a simple representative population. While association 1 has the most  ``equivalent" O7V stars, association 2 actually has a greater total O star population. This can be understood as follows: association 1 is younger than association 2, and thus contains more massive O stars with stronger ionizing fluxes than association 2; association 2 has evolved more than association 1, its most massive O stars having died out, and it is left with a large late-type O and B star UV radiation-producing population. This result is also readily apparent in the masses of the two associations. 

\subsection{Stochastic Effects\label{stoceffs}}

As with the interpretation of results from the Starburst99 model, we caution that stochastic effects may play a large role, largely unaccounted for in the stellar population analysis of these associations. In the VM05 calculation we have again assumed a fully populated IMF, which is most likely not the case, given the limits discussed above and tabulated in \cite{cervino}. Still, we may probe the extent to which we are overestimating the stellar populations by examining the behavior of the Salpeter IMF we have assumed thus far. We assume that the IMF is a probability distribution function such that the mass of the largest star formed in an underpopulated cluster of N stars decreases as N decreases. The expression for this expectation value of the maximum mass star, $<m_{max}>$, derived by Oey and Clarke (2005) is: \begin{equation}  <m_{max}> = M_{up} - \int_{10}^{M_{up}}\left[\int_{10}^{M}\Phi(m)\,dm\right]^{N}\,dM \end{equation} 

In this equation, $\Phi(m)$ is the IMF, integrated over mass, shown as two separate variables ($m$ and $M$) for the two separate integrals. For a single power-law Salpeter IMF and M$_{up}$ of 120 M$_{\odot}$, the expected maximum mass star produced does not approach M$_{up}$ until N stars with masses above 10 M$_{\odot}$ reaches $\sim1000$ (in actuality, the expectation value never reaches M$_{up}$ as the integrated function is asymptotic). Below this value of N stars, varying degrees of stochastic effects must be considered.  Since we have already normalized the IMF for each association using the UV-flux from B stars, we perform another simple integration to calculate the number of stars with masses initially greater than 10 M$_{\odot}$ in each association. Association 1 contains 25 stars greater than 10 M$_{\odot}$, association 2 contains 60, and association 5 contains only 9.  We numerically integrated equation 1 for N =25, 60, and 9, and for a  Salpeter IMF to arrive at  $<m_{max}>$ for each of the associations. These values are 80 M$_{\odot}$, 96 M$_{\odot}$, and 55 M$_{\odot}$ for associations 1, 2, and 5. These results indicate that our age estimates and total masses for association 5 may be over estimated -- if the maximum mass star in a population of this size is initially 55 M$_{\odot}$, and we have calculated the present-day upper mass limit to be 40 M$_{\odot}$ as well, we could just as easily be witnessing a zero-age population forming in-situ. Stochastic effects are less dramatic for associations 1 and 2, but nonetheless important.  The greater difference between the expected initial maximum mass and the present-day upper mass limit indicates directly that some evolution has occurred for associations 1 and 2. We stress that while our age estimates for association 5 are strict upper limits, we can have more confidence in our age estimates for associations 1 and 2. 
\begin{deluxetable*}{ccccccc}
  \tabletypesize{\small}
  \tablewidth{0pc}
  \tablecaption{Derived Properties of the Stellar Associations\tablenotemark{a}}
  \tablehead{\colhead{}&\colhead{}&\colhead{S99:}&\colhead{}&\colhead{VM05:}&\colhead{}&\colhead{}\\
  \colhead{H~II}& \colhead{N$_{O7V}$}& \colhead{M$_{tot}$}&\colhead{Age}&\colhead{N$_{OV}$}&\colhead{ M$_{tot}$}&  \colhead{Age}\\
   		 \colhead{Region}&\colhead{}&\colhead{$10^{3}$ M$_{\odot}$} &\colhead{Myrs}&\colhead{}& \colhead{$10^{3}$ M$_{\odot}$}&\colhead{Myrs}}
  \startdata
  
   1&          8.06&  1.24-1.49  &2.9 - 3.5&  11&   2.73& 4.4 - 5.0\\
   2&          5.46&  1.85-2.03 & 4.2 - 4.8& 22&    7.17&5.6 - 6.2\\
  5&           3.81& 0.61-0.84& 2.2 - 3.8&  4&    0.97&3.8 - 4.6\\
   
  \enddata
  \tablenotetext{a} {\scriptsize N$_{O7V}$ is the number of {\it{equivalent}} O7V stars in each association, calculated using the SINGG H$\alpha$ luminosity; N$_{OV}$ is the {\it{total}} number of O stars present in each association (see Section 3.2 and the appendix for details); Total stellar masses and population ages are given for two analysis methods, a comparison with Starburst99 model outputs (S99), and an independent calculation described in Section 3.2 (VM05). }
\end{deluxetable*}

\section{Discussion}

In this section we discuss the properties of the associations, clumps, and blue field objects, and compare and contrast them with those of stellar populations in the Galaxy and intergalactic H~II regions of other systems. Do stellar clusters carry an imprint of the environment in which they form, or are their properties independent of their surroundings?   We also examine the H~I properties of the environment containing the isolated associations and comment on the conditions for their formation and potential evolution. Finally, the age and mass constraints, along with physical sizes and locations, provide additional clues as to their likely fates in light of the rapid cluster dissolution inferred in many systems.  

\subsection{Comparison With Galactic OB Associations, Open Clusters, and Other Intergalactic H~II Regions}

\subsubsection{Associations}

The associations span roughly 100 pc, and contain components separated by up to 10 pc, similar to the unbound OB associations of the Milky Way.   In addition, their luminosities and masses match those of a sample of Milky Way associations within 1 kpc, and the Galactic phenomenon of similarly-sized association subgroups is directly comparable to the clumps seen here \citep{blaauw}. The properties of the associations match those of the extragalactic OB associations in seven nearby spiral galaxies with distances ranging from 6.5 - 14.5 Mpc \citep{bresolin98} equally as well. These extragalactic OB associations (diameters $\leq200$ pc) have median diameters between 70 - 100 pc, contain 6 -15 blue stars with M$_{V}\leq-4.76$,  and contain cluster candidates with V-I colors between -0.07 and 1.53. By comparison, the associations discussed here contain 3 - 9 clumps with M$_{V}\leq-3.4$, and  V-I colors between -0.90 and 0.2. Previous studies of spiral galaxy OB associations have noted a constant mean size of OB associations near 80 pc, a size that may represent the smallest step in the hierarchical structure of giant star-forming regions \citep{ein87, efremov95, ivanov96}. The sizes of these intergalactic OB associations are consistent with the mean sizes of both Milky Way and extragalactic OB associations.
 
 As noted in the Introduction, there is a growing number of objects in the literature termed ``intergalactic H~II regions,'' with photometric properties that span a wide range, and environments that range from compact groups of galaxies to tidal streams from interacting galaxy pairs. One population of four intergalactic H~II regions lies in the H~I tail to the east of NGC 7319, a member of Stephan's Quintet, a Hickson Compact Group composed of 3 interacting galaxies at 80 Mpc \citep{mendes04}. The four intergalactic H~II regions in Stephan's Quintet appear to represent brighter, more massive, and larger versions of the intergalactic H~II regions of NGC 1533. Their absolute $B$ band magnitudes cluster near -12, at the level of a faint dwarf galaxy, their derived masses are near 3$\times$ 10$^{4}$ M$_{\odot}$,  their physical sizes (FWHMs) are between 400 - 600 pc,  and their H$\alpha$ luminosities are between 10$^{38}$ and 10$^{39}$ ergs s$^{-1}$.  In contrast, the  H$\alpha$ luminosities of the NGC 1533 intergalactic H~II regions are between 10$^{37}$ and 10$^{38}$ ergs s$^{-1}$, with physical sizes $<$ 100 pc, and masses on the order of 10$^{3}$ M$_{\odot}$.

In the ring-like H~I structure surrounding the galaxy NGC 5291 (d = 62 Mpc), 29 intergalactic H~II regions have been discovered, and appear to be characterized by a ``UV-excess" compared to their H$\alpha$ emission \citep{boquien07}. The ratio of the H$\alpha$ star formation rate (SFR) to the NUV SFR (SFR(H$\alpha$)/SFR(NUV) $\sim$ 0.5 for their entire sample) is considerably lower than typical ratios for Galactic H~II regions (1 - 3). Here, we calculate SFR(H$\alpha$)/SFR(NUV) for our three intergalactic H~II regions, transforming our F250W UV fluxes (2715\AA) to NUV fluxes (2271\AA) by assuming a UV spectral slope typical of starbursting systems, F$_{\lambda}\sim\lambda^{-2.5}$, thus multiplying the UV fluxes given in Table 2 by a factor of 1.56 \citep{meurerheck99}. We use the equations presented in \cite{boquien07} for the SFRs (converted to cgs): SFR(UV) = [NUV] = 2.41 $\times$ 10$^{-40}$ L$_{NUV}$ [ergs s$^{-1}$\AA$^{-1}$] M$_{\odot}$ yr$^{-1}$; and SFR(H$\alpha$) = [H$\alpha$] = 7.9 $\times$ 10$^{-42}$ L$_{H\alpha}$ [ergs s$^{-1}$] M$_{\odot}$ yr$^{-1}$. Associations 1, 2 and 5 have [H$\alpha$]/[NUV]  of 3.0, 1.3, and 3.1, respectively, and thus do not show any UV excess as observed for the intergalactic H~II regions around NGC 5291. \cite{boquien07} explain the UV-excess in their sample as the  result of a quasi-instantaneous burst of star formation, resulting in two populations of stars, one roughly 4 Myrs old, and the other 10 Myrs. For comparison, their PSF in the NUV has a characteristic width of  5$\arcsec$, 75 times larger than our PSF.  At 62 Mpc, 5$\arcsec$ covers 1.5 kpc, and would therefore encompass most of the HRC field of view at the distance to NGC 1533, and quite possibly all of the visible star formation around regions 1 and 2, which we show spans a range of  possible ages (2-9 Myrs).

\subsubsection{Clumps and Blue Field Objects}

Aside from single O and B stars, the closest Galactic analogs to the clumps and blue field objects are young open clusters. For example, \cite{bonatto06a} presents a detailed analysis of the properties of NGC 6611 (the Eagle Nebula) deriving an age of 1.3$\pm$0.3 Myr, a limiting radius of 6.5$\pm$0.5 pc, and an observed mass of 1600 M$_{\odot}$.  Viewed from a distance of 19 Mpc, a young open cluster such as NGC 6611 would appear unresolved in the HRC and similar to one of the clumps or blue field objects. Moreover, the integrated photometric characteristics of Galactic open star clusters younger than 10 Myrs are fairly consistent with those of the clumps and field objects. Of the 140 open clusters (masses between 25 and 4 $\times$ 10$^{4}$ M$_{\odot}$) tabulated in \cite{lata02}, 17 are younger than 10 Myrs, with absolute V band magnitudes between -4.5 and -9, and V-I colors (for only 6 of the 17) between 0 and -0.27. These colors are similar to those found for the clumps and blue field objects listed in Tables 2, 3, and 4. However, the properties of the clumps and blue field objects are also in line with single O and B stars, seen in Figures 5 and 6. In the end, we cannot distinguish between single stars and young open clusters at the physical resolution limit of the HRC. 

The clumps and blue field objects also have extragalactic analogs, and more extreme counterparts.  The photometric properties and sizes of the clumps and blue field objects  align with those of the faint 1-10 Myr clusters in the LMC and SMC studied by \cite{hunter03}.  Compared to the young clusters in the disk of M82, a violently starbursting system, the clumps and blue field objects have similar V-I colors (on the blue side) and absolute $V$-band magnitudes. The range of V-I colors for the M82 young clusters is -0.05 to 2.34, with absolute $V$-band magnitudes ranging from -11.2 to -5.3, having a mean  of -6.75 \citep{kons07}. However, the young massive ($\sim$ 10$^{5}$ M$_{\odot}$) clusters of the Antennae galaxies exhibit a range of properties nearly on a different scale from those of the stellar populations powering NGC 1533's intergalactic H~II regions. The clumps lie on the faintest end of the Antennae cluster luminosity function constructed by \cite{whitmore99}, contain far fewer stars (indeed, they may be single stars), and for the most part, occupy the bluest end of the V-I color distribution.  Overall, the properties of the clumps and blue field objects show no deviation from the continuum of properties for known star clusters (if they are not single stars) in both our Galaxy and others. 

\subsection{Gas Properties} 

 The H~I distribution shown in Figure 1 reveals a ring around NGC 1533
 with a total H~I mass of $6\times10^{9}$ M$_{\odot}$
 \citep{emma03}. The southeast cloud, the site of the star formation
 discussed throughout this paper, contains approximately 1/3 of this
 total H~I mass ($2.4\times10^{9}$ M$_{\odot}$), and has a maximum H~I
 column density of $4 \times10^{20}$ cm$^{-2}$. The three associations
 do not lie in the densest regions of the SE H~I cloud, but rather in
 regions well below the canonical neutral gas surface density cutoff
 for star formation on global scales ($10^{21}$ cm$^{-2}$,
 \nocite{skillman87} Skillman 1987). Of course, the Q scenario
 predicts that this critical gas density threshold ($\Sigma_{crit}$)
 scales with velocity and radius \citep{kennicutt89}. For a flat rotation curve and parameters fitted in \cite{kennicutt89}, $\Sigma_{crit}$ (M$_{\odot}$ pc$^{-2}$) = 0.4 $\frac{c}{6}$$\frac{V}{R}$, where c is velocity dispersion and V is the rotation velocity of the galaxy,
 both in km s$^{-1}$, and R is the projected distance from the galaxy
 nucleus in kpc. Given a fitted rotation velocity of the gas around
 NGC 1533 of 150 km s$^{-1}$ and a velocity dispersion of 10-30 km
 s$^{-1}$ \citep{emmaconf}, at the projected radii of the isolated H~II
 regions (19 - 31 kpc), the threshold column densities of the H~I would
 range between $4 - 20 \times10^{20}$ cm$^{-2}$. The observed column
 densities at the positions of the intergalactic H~II regions appear to fall
 below this range in critical surface density (see Table 1);  although it is possible 
 the true H~I column densities are
 underestimated due to the 1 arcmin beam.
 It should also be noted that though the H~I ring appears to be rotating
 around NGC 1533, it is not part of a smoothly rotating disk
 as studied by \cite{kennicutt89}.
 
 Across entire galaxies, star formation regions tend to align with H~I
 peaks, but this global correlation breaks down on
 sub-kiloparsec scales \citep{kennicutt07}.  On a local scale,
 low-levels of star formation can proceed despite lower than normal
 neutral gas column densities.  Similar to the H~II regions outside
 NGC 1533, the Stephan's Quintet intergalactic H~II regions are not
 associated with H~I peaks in the tidal tail (at levels below
 10$^{20}$ cm$^{-2}$) \citep{mendes04}. Additionally, the H~II region
 found in the Virgo Cluster by \cite{gerhard} lies in no detectable
 amount of H~I gas (down to $10^{19}$cm$^{-2}$; \nocite{hioosterloo}
 Oosterloo \& Van Gorkom 2005).  There must be a reservoir
 of yet undetected molecular material at the positions of these
 intergalactic H~II regions.   The Atacama Large Millimeter Array (ALMA)
 will shed light on the molecular conditions required to trigger
 star formation outside the disk of galaxies.

\begin{figure}
\begin{center}
\plotone{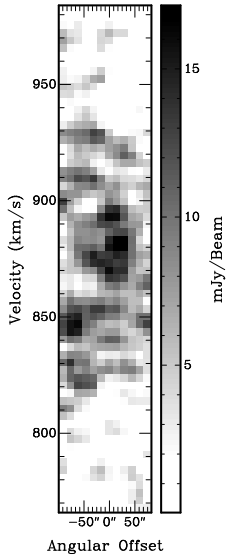}
\end{center}
\figcaption{A slice in position-velocity space across the long diameter of the apparent H~I concentration just to the north of associations 1 and 2.   This figure shows that the concentration is made up of numerous clouds across a wide range of velocities.  The noise level is at $\sim4$ mJy/bm and the beam is 1 arcminute \citep{emma03}. }

\end{figure}

 Though it may appear as though associations 1 and 2 lie near a relatively dense
 cloud of H~I (just to the north of the associations), there are actually multiple clouds along the
 line of sight, and not a single, rotating cloud. Figure 8 shows the lack of a smooth velocity gradient at the position of this apparent H~I concentration in a position velocity slice across the long diameter of the cloud.  The total H~I mass of the
concentration of gas clouds in the vicinity of associations 1 and 2 
is $3\times10^{8}$ M$_{\odot}$, similar to the amount of H~I found in the vicinity of the potential tidal dwarf galaxies
of the Antennae \citep{hibbard01} and NGC 5291 \citep{bd06}, but lower
than what is typically found for tidal dwarf galaxies ($\geq10^{9}$ M$_{\odot}$ e.g. Duc \& Mirabel 1994, 1998; Hibbard \& Barnes 2004; \nocite{ducmirabel94,ducmirabel98,hibbardbarnes04}).
The velocity dispersions in the vicinity of NGC 1533's intergalactic H~II regions are near 30 km
 s$^{-1}$, with gradients in the range 7-50 km s$^{-1}$
 kpc$^{-1}$ \citep{emmaconf}.   The relatively high velocity
 dispersions may lend some clue to the conditions of their
 formation, as cloud-cloud collisions are known to trigger star
 formation.  These velocity dispersions are 
 much higher than the H~I velocity dispersions 
 measured in the vicinity of the tidal dwarf galaxy candidates in the
 Antennae (5-15 km s$^{-1}$; \cite{hibbard01}), and the overall kinematic structure is much
 more irregular than that of both the Antennae and other potential tidal dwarf galaxies (e.g., Bournaud et al. 2007).  We plan to explore the relation between H~I column
 densities and velocity dispersions for star formation in the far
 outskirts of galaxies in a future paper including a larger sample of
 intergalactic H~II regions and high-resolution H~I data.

 While the stellar masses revealed here do not place
 the star formation in the HI ring in the category of tidal dwarf galaxy (TDG) candidates, the
 location of associations 1 and 2 in the tip of the SE H~I tidal tail
 may actually be a prime one for survival.  In a set of 96 N-body
 simulations, \cite{bd06} found that of all structures formed during
 mergers, the only objects that are long-lived are those formed in the
 extremities of the tidal tails. The tips of tails are often sites of
 kinematical pilings-up of gas \citep{duc04}, contain populations less
 prone to fall back upon the progenitor galaxy, and lie far beyond the
 reach of disruptive galactic tidal forces \citep{bd06}. However,
 these simulations were limited to stellar tidal structures with
 masses greater than 10$^{8}$ M$_{\odot}$, a criterion our
 associations surely do not meet.   If one assumes the gas
 in the clump near associations 1 and 2 is part of a bound structure,
 the resulting dynamical mass is 20-25 times higher than the baryonic
 mass in this region.  This abnormally high dynamical mass for a TDG, together with the structure of the clouds, indicate the stellar associations are unlikely to evolve into a relatively long-lived tidal dwarf galaxy.

\subsection{Fate and Evolution}

	In the recent literature, there has been much focus on the age distribution of star clusters as a way of examining their rate and nature of dissolution.  We cannot construct an age distribution of the 19 individual clumps and 18 field objects  because they all have $V$-band absolute magnitudes considerably fainter than -9, a limit indicating that single stars dominate their output radiation.  We can, however, comment on the associations and their relationship to the clumps and blue field objects in the context of cluster disassociation.  In this section, we assess the survival of the associations by considering the locations of the objects in the HRC fields relative to each other, their relative sizes, and age and mass constraints from stellar evolution models.

\subsubsection{Rapid Dissolution} 

Star clusters of a wide range of masses and in various environments are subject to a large number of possible internal and external evolutionary effects. In the Milky Way, embedded protoclusters less massive and more densely concentrated than the isolated associations discussed here are observed to dissolve on fast timescales (``infant mortality"), with less than 4-7\% surviving as bound clusters \citep{ladalada}. In the extreme environment of the colliding Antennae galaxies, the survival prospect is just as grim for more massive young ``super star" clusters \citep{fcw}. In more quiescent systems like the SMC, ``infant mortality" may play a much smaller role, if any; however, its role has not been fully investigated for clusters younger than $\sim$10 Myrs due to a detection bias against clusters having nebular emission in the SMC samples \citep{gieles07,degrijs07}.  
The fate of the stellar systems powering intergalactic H~II regions (ages $< 10$ Myrs; masses 10$^{3}$ - 10$^{4}$~M$_{\odot}$) has not yet been explored in the literature. On one hand, their fate may be similar to that of Galactic OB associations, thought to expand freely for tens of Myrs \citep{blaauw,debruijne99}. In this case, one of their most massive clumps (subgroups) on average ends up as a bound, open cluster such as the Pleiades \citep{ladalada}. On the other hand, their location in an environment vastly different from that of the Galactic disk may result in a different fate. Furthermore, objects of similar masses and sizes to these isolated associations seem to be able to survive for hundreds of Myrs in the SMC \citep{degrijs07}, while the vast majority appear to dissolve on the order of tens of  Myrs in the Antennae Galaxies \citep{fcw}.  In this context, it is relevant to explore what will become of these isolated associations given our high-resolution photometric data.

 The scenario of rapid cluster dispersion requires some disruptive effect other than the general stellar evolutionary processes and gravitational effects such as stellar evaporation driven by two-body relaxation considered in typical cluster fading laws.
 Most commonly proposed is that the energy and momentum input from massive stars (in the form of winds, jets, ionizing radiation, and supernovae) remove the cluster's ISM, leaving it unbound and freely expanding \citep{hills80,goodwin97,fcw}. These effects result in an average expansion rate of the stars in Galactic OB associations of 5 km s$^{-1}$, or 5 pc in 1 Myr, estimated analytically \citep{hills80}. Observationally, the direct measurement of an expansion speed will have to wait until the launch of  Gaia. However, under realistic assumptions of expansion, the stars' proper motions and radial velocities (along with the best positions available from Hipparcos) are consistent with linear expansion coefficients of about 5 km s$^{-1}$ per 100 pc \citep{blaauw,bobylev07}. Thus, typical Galactic associations can expand for only tens of Myrs before dispersing into the general field population.  If these intergalactic OB associations expand similarly, they eventually will not be recognizable as distinct groups of stars. The expansion of associations also implies that larger associations, will be, on average, older than more compact associations. Of the three studied here, Association 5 is the youngest (2 - 4.5 Myrs), and also has the smallest visible physical extent of roughly 60 pc. On the other hand, association 2 is the oldest (4.5 - 6.5 Myrs), and has the largest physical extent of roughly 100 pc.  This age-size sequence is consistent with a scenario of dispersion.  Association 5 may still be recognizable as an association of massive stars for some time, whereas associations 1 and 2 likely have less time before they will be indistinguishable from the blue field objects.   
 
Do the blue field objects then represent stars and/or star clusters that were once physically linked with the associations and have since dispersed into the field?  Given the typical expansion rate of OB associations, if the blue field objects were once part of association 1 or 2, they would have to be 30~-~240 Myrs old to account for the observed separations of 150~-~1200 pc.  The colors of the blue field objects are not consistent with this range of ages.  The single stars dominating the radiation output of the blue field objects are roughly consistent with main sequence stars ranging from 8 M$_{\odot}$  to 40 M$_{\odot}$ (consistent with B2V - O5V stars; see Figures 5 and 6).  Geneva stellar tracks \citep{lejeune} predict that an 8 M$_{\odot}$ star has a V-I color below zero as it burns hydrogen, and reddens by over 1 magnitude at an age of 27 Myrs, while a  40 M$_{\odot}$ star has a V-I color below zero as it burns hydrogen, and reddens by more than 1 magnitude near 6 Myrs.   Since the blue field objects generally have V-I colors below zero, they are unlikely to be older than 6 - 27 Myrs, and therefore would not have had time to reach their current projected separations from the nearest association (1 or 2) at the assumed velocity of 5 km s$^{-1}$.    There is also no indication that the field objects further from the association are redder (and, by inference, older) than those that are nearby. Some of the blue field objects may also represent single runaway O or B stars ejected from the associations at maximum speeds ranging from 200 - 350 km s$^{-1}$ \citep{Hoogerwerf01,leonard90}. This speed would give the stars the ability to travel between 1200 - 9450 pc in their estimated lifetimes, but this phenomenon is highly unlikely to account for all 18 blue field objects, if any at all.

A more likely scenario for the blue field objects in the context of cluster dissolution is that some of them once formed their own association that is no longer easily recognizable as such. In the Galaxy, the Cassiopeia-Taurus group may represent a similar grouping of objects with a common origin in a more advanced state of disintegration \citep{blaauw,deZeeuw99}. Here, F5, F6, F8, F9, and F11 (and other tiny sources visible in the same locale, but undetected by SExtractor, see Figures 2 and 4) appear to be clustered together themselves, in an ostensibly  young, diffusing complex with a total extent of about 150 pc and average object projected separations of 20 pc.  The radius of this potential complex, given average association expansion velocities and assuming all its components formed at roughly the same position, is indicative of an age of $\sim20$ Myrs, which is roughly consistent with the range of ages possible for the blue field objects (see above).  In the SINGG H$\alpha$ image (1.37$\arcsec$ resolution), we do not detect unresolved H$\alpha$ emission at the location of this ``complex" as we do for associations 1, 2, and 5.  Assuming the H~II region potentially associated with this diffusing complex is similarly unresolved in the SINGG image (likely, given the 1$\arcsec$ FWHM of the complex of objects), the upper limit to the H$\alpha$ flux at the location of this ``complex" is 9.96$\times10^{-17}$ ergs s$^{-1}$ cm$^{-2}$, based on 5$\sigma$ point source detection limits of the SINGG data.  Dividing this upper limit by the sum of the UV flux of F5, F6, F8, F9 and F11, we derive an upper limit Log~$(F_{H\alpha}/F_{2715\AA})$ of 1.44. The subsequent lower age limit of the diffusing ``complex" based on Figure 7 is 5.6 Myrs, at least 1-2 Myrs older than the ages of associations 1, 2 and 5 based on this same plot, and consistent with the idea that these blue field objects represent an association in the process of dissolving while the other associations are comparably more intact. 

One can also consider what should be observable in the field if we assume a continuous and homogeneous low rate of star formation in this region for the past 100 Myrs.  In this scenario, if we see 2 associations of 2000 M$_{\odot}$ that are under 10 Myrs old, then it follows that we would expect 2 associations of 2000 M$_{\odot}$ every 10 Myrs  for the last 100 Myrs, or twenty 2000 M$_{\odot}$ associations in this region.   Independent of the processes responsible for rapid cluster dissolution, we should consider the effects of fading due to mass loss from stellar evolution and tidal effects, as this will decrease the number of clusters we expect to see.  Because our detection limits are similar to those of \cite{hunter03} in their study of the Large and Small Magellanic Clouds, we assume a cluster fading function similar to theirs, where M$_{fade}$~=~982(t/Gyr)$^{0.69}$~M$_{\odot}$. Thus, without rapid cluster dissolution, we would be able to detect similar mass associations with ages up to $\sim$1 Gyr (approximately the age of the HI ring, Ryan-Weber et al. 2003a \nocite{emmaconf}).  That we do not detect associations independent of those directly associated with the H~II regions indicates that the clusters are either rapidly dispersing or this is the first generation of star formation in the HI stream.   

In the scenario of rapid dissolution, we may still expect to find background fluctuations in our images representative of previous star formation in the HI stream that has subsequently dispersed. In our HRC data, we see no evidence of an underlying older population after investigating the noise distributions in each band. The shape of the noise distribution is nearly perfectly Gaussian in all cases, and in UV, V, and I we probe down to 28.0, 28.9, and 29.0 AB magnitudes, respectively. While this does not rule out an older population of dispersed clusters (or a previous stellar component to the HI stream), if one exists, it is below the noise level.  Considering an expansion rate of 5 km s$^{-1}$, we can estimate the expected cluster fading due to its dissociation alone. A linearly increasing cluster radius with time \citep{blaauw,fcw}, results in the surface brightness decreasing with the square of time (in fact, the surface brightness will decrease faster because as stars age and die, the cluster becomes fainter). For areas the size of these associations ($\sim$186 pixels$^{2}$), our HRC images give a 5$\sigma$ surface brightness limit of 24.3 mag arcsec$^{-2}$ in the $V$ band. We calculate the $V$-band surface brightnesses of the associations 1, 2, and 5 (using their half-light radii) to be 21.85, 22.26, and 22.0 mag arcsec$^{-2}$, respectively.  Thus, in 14 Myrs, when the average radius of the associations has increased by a factor of 2.5 (from 47 pc to 117.5 pc), the surface brightness of the associations will fade below the detection limit of the HRC images. It is consistent with the idea of rapid cluster dissolution that we see one nearly dissolved association roughly 6 - 20 Myrs old, and several younger, more intact associations.

\subsubsection{Alternative Scenario: A First Generation of Stars}

Contrary to the above argument for rapid dissolution, the associations, clumps, and blue field objects may represent a distribution of a first-generation of star formation in which the stars and clusters have not strayed far from their location of formation. We emphasize that this scenario is self-consistent only if star formation in this region began less than 10~-~15 Myrs ago, otherwise we would see older, intact associations similar in mass to those observed.  In this scenario, the blue field objects would occupy the low mass end of the initial cluster mass function (CMF) rather than represent the remnants of dissolved larger stellar associations. That the youngest association (association 5) is the most compact may simply indicate a lower initial mass of this association compared to the others, and not that it is part of an age-size sequence of cluster dissolution.  If the observed associations comprise a first generation of stars, then we cannot use these HRC data to untangle their fates. 

In this regard, the isolated and singular association 5 may be most likely to be an initial episode of star formation.  For, if star formation has been continuous in this region for any period of time greater than 10 Myrs, the absence of any sign of dispersed cluster remnants in its vicinity may be inconsistent with cluster disintegration. That this youngest association remains isolated in comparison with its older counterparts offers some indication that it represents an initial episode of star formation in this northern region of NGC1533's H~I ring. There are a few related caveats to consider: 1) even if association 5 represents a first generation of star formation, rapid dissolution may subsequently proceed 2) association 5 could simply represent the youngest, most compact component of the age-size sequence we expect for cluster dissolution 3) the star formation in this northern region may have begun later than that in the southeast, and so should simply be considered separately. 



\section{Summary and Conclusion}

HST ACS/HRC images in UV, V and I bands resolve the three very young stellar associations powering the intergalactic H~II regions that lie up to 30 kpc from the stellar disk of the S0 galaxy NGC 1533. We have catalogued and performed circular aperture photometry on 48 individual sources in the HRC images, 27 of which appear to be associated with the ongoing star formation. From background galaxy statistics and broadband colors, we have determined that the other 21 sources are likely distant background galaxies. Furthermore, we have defined a range of sources in the images: the term ``associations" refers to the total stellar populations powering the H~II regions; the term ``clumps" refers to the individual resolved sources composing the associations; and the term ``field objects" refers to the individual sources detected separately  from the associations. Using two different methods,  (Starburst99; \nocite{starburst99}Leitherer et al. 1999, and an independent calculation derived from \nocite{vacca94, martins} Vacca 1994 and Martins 2005, see appendix) in combination with our broadband colors and magnitudes, and H$\alpha$ luminosities from SINGG \citep{SINGG}, we have placed upper limits on the the total numbers of stars, masses and ages of each association. The masses of the isolated associations range from 600 M$_{\odot}$ to 7000 M$_{\odot}$, and the ages range from 2 - 6 Myrs. Given the faint luminosities of the stellar associations, we have also considered the importance of stochastic effects upon our calculations. Colors and absolute magnitudes of many of the clumps within the associations and field objects are consistent with single O and B stars dominating the output radiation. 

There are three results from our analysis that deserve emphasis: 
\begin{itemize}
\item{Despite their location in an environment that is vastly different from that of a Galactic spiral arm, the photometric properties and sizes of the associations, clumps, and blue field objects are consistent with those of Galactic stellar associations and clusters and/or stars. This result suggests something uniform about the star-formation process on scales of tens of parsecs, independent of environment.}
\item{The level of star formation in the SE H~I cloud of NGC 1533 is lower than that observed in other systems (e.g. NGC 5291; \nocite{boquien07} Boquien et al. 2007).  Furthermore, low H~I column densities (10$^{20}$cm$^{-2}$) and high H~I velocity dispersions ($\sim30$km s$^{-1}$) characterize its environment. These data hint that there may be some relation between SFR, H~I column density and velocity dispersion that holds outside of massive gas disks, on local scales. These data are also consistent with the recent finding that H~I surface density and star formation rate densities do not correlate on sub-kiloparsec scales \citep{kennicutt07}.  The kinematics of the H~I in the vicinity of the H~II regions do not suggest the stellar associations will evolve into a larger tidal dwarf galaxy.}
\item{The stellar associations detected here are either in the process of rapid dissolution or represent the first generation of stars in the gaseous ring around NGC 1533.  If the star formation in the SE part of the H~I ring has been continuous over the last 20 Myrs to 1 Gyr, we are directly witnessing the rapid dissolution of young clusters and associations. Not only do the associations display a range of ages and sizes consistent with cluster dissolution, but the age constraints and locations of several clustered blue field objects indicate they are the last visible remnants of an already dispersed association. 
If instead, these isolated associations and blue field objects represent a first generation of stars in the stripped H~I of NGC 1533, we cannot determine the ultimate fate of the associations. This possibility of a first generation of stars is then uniquely suited for exploring star formation triggers and initial conditions without the influence of stellar feedback.   Given that their Galactic OB association analogs are generally thought to be gravitationally unbound and freely expanding \citep{blaauw,hills80}, along with evidence for star cluster ``infant mortality'' in various clustered stellar systems (e.g. \nocite{ladalada, fcw} Lada \& Lada 2003; Fall et al. 2005), we find it most likely that the stellar populations of the intergalactic H~II regions will soon ($< 20$ Myrs or so) no longer be detectable nor recognizable as associations of stars; rather, they will be building a background population of young objects in the halo of NGC 1533. }

\end{itemize}

\acknowledgements 
 These observations were supported through NASA grant HST-GO-10438. JKW acknowledges support from an NSF Graduate Research Fellowship. GRM acknowledges additional support from NAG5-13083 (the LTSA program). JKW gratefully acknowledges very helpful email conversations with Bill Vacca, the generosity of Iraklis Konstantopoulos with M82 cluster photometry,  and John Blakeslee with providing the revised distance measurement to NGC 1533. This research has made use of the NASA/IPAC Extragalactic Database (NED) which is operated by the Jet Propulsion Laboratory, California Institute of Technology, under contract with the National Aeronautics and Space Administration.

\appendix
\section{Independent Calculation of Cluster Ages and Masses}
Here we present the details of the estimate of the total number of ionizing stars present in each star-forming region derived from \cite{vacca94}. In Section 3.2, we use this method to determine association ages and masses, and compare the results with those of Starburst99. This method assumes no stellar evolution off the zero age main sequence, but is still provides reasonable estimates for the total number of main sequence O stars because the emission rate of ionizing photons does not vary substantially as massive stars evolve \citep{vacca94}.  In A.1, we outline the calculations of the total number of O stars based on the observed ionizing flux of each association. In A.2, we use the observed UV fluxes to constrain the number of B stars, set the upper mass limit to the present-day mass function, and normalize the IMF for each isolated association. 

\subsection{The Number of O Stars}

An H$\alpha$ luminosity corresponds to an emission rate of ionizing photons, Q$_{0} = 7.33\times10^{11}$ L$_{H\alpha}$  s$^{-1}$ with L$_{H\alpha}$ in ergs s$^{-1}$ cm$^{-2}$ \AA$^{-1}$  \citep{osterbrock}.  Using this conversion factor, one can derive the number of  ``equivalent" stars of a given subtype powering an H~II region. This number of  ``equivalent" ionizing stars, while a simple characterization of the stellar population, does not represent the {\it{total}} number of ionizing stars. With this notion in mind, \cite{vacca94} present values for the conversion factor ($\eta_{0}\equiv$ N$^{\prime}_{O7V}$/N$_{OV}$) between the ``equivalent'' number of O stars of a given subtype and the full ionizing population, utilizing the most current stellar atmosphere and stellar evolution models available in 1994. In this appendix, we describe our calculation of the total ionizing population of each H~II region. Essentially, our method follows the calculations outlined in \cite{vacca94}, but uses the recently updated T$_{eff}$ - spectral type calibration of \cite{martins}.  \cite{vacca94} thoroughly details the calculation of this conversion factor, $\eta_{0}$, and we refer the reader to that paper for a full description of the calculation. 

\begin{figure}
\figcaption{ On the left: The ratio of the ionizing photon emission rates in photons s$^{-1}$ as a function of  stellar mass between two sets of stellar parameters, those presented in \cite{vacca96} and \cite{martins}. The separate functions Q$_{0}$(M) are quite similar, yet the differences produce significant changes in the conversion factors used to determine the total ionizing stellar populations (see Table 4 and the appendix for values and a detailed description).  Below 16 M$_{\odot}$, the two functions are exactly the same. On the right: The polynomial fits to the data points of \cite{martins} are plotted. The masses used to generate Q$_{0}$(M) from \cite{vacca96} were the evolutionary masses, while the masses used to generate Q$_{0}$(M) from \cite{martins} were the spectroscopic masses (the only masses tabulated). Using the spectroscopic masses does introduce some error to our calculation, though slightly tempered by our use of only those masses below 60 M$_{\odot}$, and the lower effective temperature scales of \cite{martins}.  \label{massqo}}
\begin{center}
\plotone{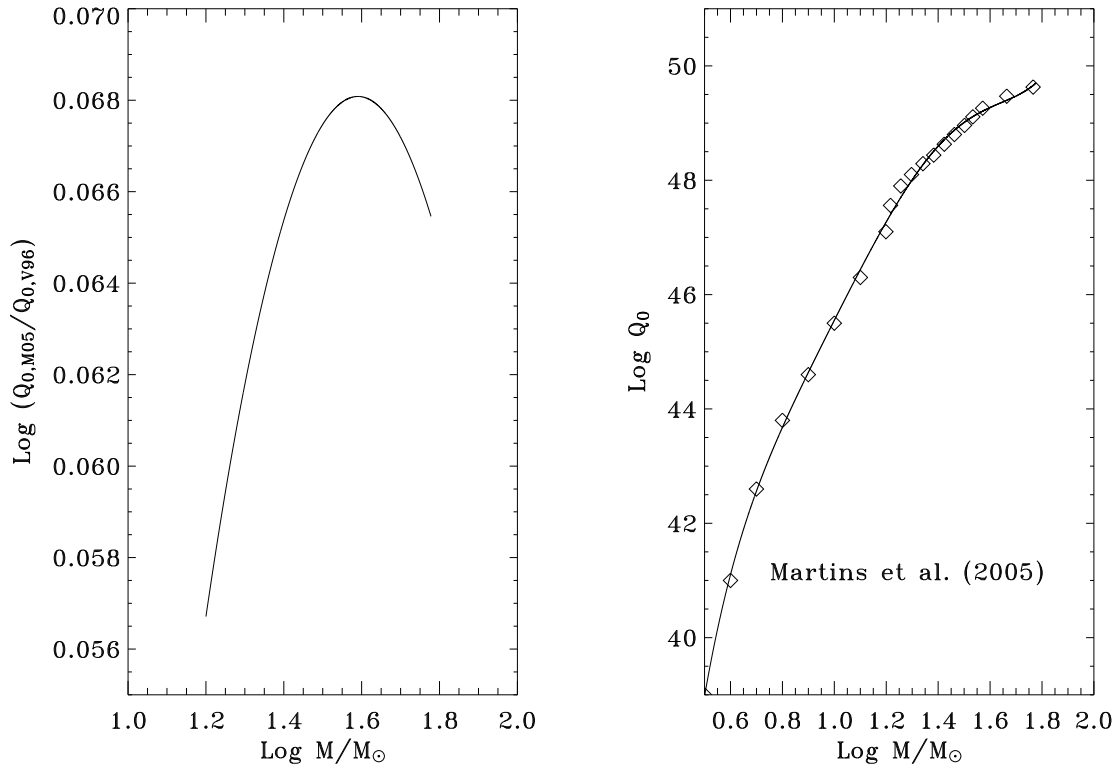}
\end{center}
\end{figure}

The motivation for recalibrating $\eta_{0}$ using \cite{martins} arose from the significant difference in their stellar parameters from previous results. Taking non-LTE, and wind and line-blanketing effects into account leads to  effective temperature scales that are cooler than previously reported scales  by 2000 to 8000K. Furthermore, the inclusion of metals in the model atmospheres of \cite{martins} modifies the ionizing fluxes of the O stars.   A true detailed recalculation of $\eta_{0}$ would have to include the new evolutionary tracks and regenerate stellar models, a task beyond the scope of this paper. Instead, we present a ``re-estimation" of the conversion factor $\eta_{0}$ that hinges upon the number of ionizing photons produced
as a function of stellar mass, Q$_{o}$(M). Figure \ref{massqo} shows this function plotted versus stellar mass for two sets of stellar parameters taken from tables presented in \cite{vacca96} and \cite{martins}. Below 15.8 M$_{\odot}$, we used ionizing fluxes read from figure 1 in \cite{vacca94} for both plots, making our two power-law fits identical in this regime. This approximation is valid since 1) the lower mass stars contribute very little to the ionizing radiation and 2) the two T$_{eff}$ - spectral type calibrations are roughly the same below about 16 M$_{\odot}$. The solid lines represent the broken power-law fits (a series of linear fits in log-log space) to the data.  

 We begin our re-estimation of $\eta_{o}$ by considering the equation for the conversion factor presented in \cite{vacca94} and applying the values tabulated in \cite{martins}. From physical definitions, \cite{vacca94} derives the following equation: \begin{equation} \eta_{o} = \frac{\int_{M_{low}}^{M_{up}} \Phi(M) Q_{0}^{ZAMS}(M) \,dM}{Q_{0}^{O7V}\int_{M_{OV}}^{M_{up}} \Phi(M) \,dM} \equiv  N^{\prime}_{O7V}/N_{OV}\end{equation} Here, $\Phi(M)$ is the present day mass function with the form AM$^{-(1+ x)}$, where A is the normalization constant and x is the power law index (x=1.35 for a Salpeter mass function). Q$_{0}^{ZAMS}$(M) is the Lyman-continuum emission rate in photons s$^{-1}$ as a function of stellar mass for stars on the main sequence. Using the relation Q$_{0} = 7.33\times10^{11} L_{H\alpha}$ s$^{-1}$ \citep{osterbrock}, the H$\alpha$ luminosities given in Table 1, and the fiducial Q$_{0}^{O7V}= 48.63$ \citep{martins}, we calculate the number of equivalent O7V stars ( N$^{\prime}_{O7V}$) in associations 1, 2 and 5 to be 8.06, 5.46, and 3.81, respectively. We then evaluate Equation A1 to calculate $\eta_{0}$. From the definition $\eta_{0}\equiv$ N$^{\prime}_{O7V}$/N$_{OV}$, we calculate the total ionizing main sequence O star population (N$_{OV}$) for a range of upper mass limits (M$_{up}$) given Q$_{o}^{ZAMS}$(M) from Figure \ref{massqo}b. The lower mass (M$_{low}$) is taken to be 3 M$_{\odot}$ \citep{schmidt-kaler}, and the lower mass limit to O stars (M$_{OV}$) is taken to be 15.8 M$_{\odot}$ \citep{martins}. The total numbers of main sequence O stars in each association are estimated at 11, 22, and 4 for associations 1, 2 and 5 using the present-day mass function upper mass limits we determine in the following section. 
 
Table 6 gives three different calculations of $\eta_{0}$ for solar metallicity, a Salpeter IMF, and  various values of M$_{up}$. The first column of $\eta_{0}$ is taken directly from \cite{vacca94}, in which $\eta_{0}$ was calculated at larger mass intervals. For comparison, we also give $\eta_{0}^{\prime}$ for the Q$_{o}$(M) of \cite{vacca94} calculated by our method of function fitting, integrating, and our choice of a lower mass for O stars of 15.8 M$_{\odot}$. $\eta_{0}^{\prime\prime}$ is this re-estimation of Vacca's original conversion factor using the recent stellar models of \cite{martins}. Our values of $\eta_{0}^{\prime}$ for the model of \cite{vacca94} agree well with the values of $\eta_{0}$ given directly in \cite{vacca94}. Furthermore, it is readily apparent that the most recent T$_{eff}$-spectral type calibration of \cite{martins} has impacted the values of $\eta_{0}$ quite significantly. 

We caution that these values of $\eta_{o}$ should not be used as more than estimates, because of the assumptions and approximations used in the calculations. Still, these are the most up-to-date estimates of their kind. \cite{martins} quote errors in their masses between 35 and 50\%, which results in $\eta_{0}$ having 15-20\% errors. While this calculation technically requires the use of an evolutionary (model dependent) mass, Martins et al. give only the theoretical spectroscopic (M$_{spec} = \frac{gR^{2}}{G}$ ) masses for each O star spectral type. Finally, the lower mass limit to O stars is quite poorly defined, which adds at least 10\% to the error in the calculation of $\eta_{0}$. The discrepancy between evolutionary and spectroscopic masses is a known issue with stellar models, but is somewhat alleviated by the decrease in the effective temperature scale of \cite{martins}. We also stress that we have done this analysis for only main sequence solar-metallicity O stars. Varying the metallicities of the modeled stars will also have a small effect on the outcome. 

\subsection{The Number of B Stars}

We calculate the total non-ionizing B star population expected from the UV luminosity of each association taking a path similar to that taken above.  If the total UV luminosity is produced by only O and B spectral types (to a good approximation), and we can quantify the ionizing population by the H$\alpha$ luminosity, then we can also quantify the population of stars that contributes exclusively to the UV luminosity.  Given the number of {\it{equivalent}} O7V stars from the H$\alpha$ luminosity, we calculate the number of {\it{equivalent}} B9V stars from the UV luminosity as follows: \begin{equation} F_{UV} = N_{O7V}\times F^{O7V}_{UV} + N_{B9V}\times F^{B9V}_{UV} \end{equation} In this equation, $F^{O7V}_{UV}$ is F$_{\lambda}$ of an O7V star  in the F250W bandpass calculated using SYNPHOT ($6.64\times10^{-19}$ ergs s$^{-1}$ cm$^{-2}$ \AA$^{-1}$ at the distance of NGC 1533 from Kurucz 1993 spectral atlas in the SYNPHOT database), $F^{B9V}_{UV}$ is F$_{\lambda}$ of an B9V star in the F250W bandpass ($9.20\times10^{-22}$ ergs s$^{-1}$ cm${-2}$ \AA$^{-1}$ at the distance to NGC 1533), and $F_{UV}$ is the total $F_{\lambda}$ of the object measured in the HRC F250W  image. For associations 1, 2, and 5 we calculate 3000, 8200, and 1100 equivalent B9V stars respectively. 

Given the number of equivalent B9V stars for each association, an analysis similar to what we described above was undertaken to determine the {\it{total}} number of  B stars that contribute to the UV luminosity of the intergalactic H~II regions. In the same sense as for O stars, a conversion factor between a number of  ``equivalent" B stars of a certain spectral type and the full population of B stars can be determined: $\eta_{o}(B) \equiv $N$_{B9V}^{\prime}$ / N$_{BV}$. In this case, N$_{B9V}^{\prime}$ is the number of equivalent B9V stars (see above), and N$_{BV}$ is the total B star population. The corresponding integral for the calculation of the conversion factor is simply: \begin{equation} \eta_{o}(B) = \frac{\int_{M_{B9V}}^{M_{up}} \Phi(M) F_{\lambda}^{ZAMS}(M) \,dM}{F_{\lambda}^{B9V}\int_{M_{BV}}^{M_{up}} \Phi(M) \,dM} \equiv N_{B9V}^{\prime} / N_{BV}\end{equation} We generate F$_{\lambda}^{ZAMS}$(M) for  all B spectral types using the masses and temperatures of \cite{schmidt-kaler} and calculating the flux in the F250W bandpass of each spectral type using SYNPHOT and the Kurucz spectral type models. $\Phi(M)$ is the  same present-day mass function used previously. Next, we calculate the {\it{total number}} of non-ionizing B stars by integrating Equation A2 for masses between 3 and 15.8 M$_{\odot}$. We estimate a total B star population of  150, 400, and 50 in associations 1, 2, and 5, respectively.  

Finally, for comparison with our original calculation, we use the normalization of the present day mass function from this calculation, and integrate $\Phi$(M) between 15.8 M$_{\odot}$ and various values of M$_{up}$  to determine the total number of main sequence O stars. The total number of O stars calculated using the H$\alpha$ luminosity should match the total number of O stars calculated using the UV flux at some upper mass limit, interpreted as the maximum mass of the PDMF.  For isolated associations 1, 2, and 5 these upper masses are 35 M$_{\odot}$, 25 M$_{\odot}$, and 40 M$_{\odot}$. Using the values of $\eta_{0}$ presented in \cite{vacca94} gives slightly higher upper masses of 40 M$_{\odot}$, 28 M$_{\odot}$, and 50 M$_{\odot}$ for associations 1, 2, and 5, respectively.

\begin{deluxetable}{cccc}
  \tabletypesize{\small}
  \tablewidth{0pc}
  \tablecaption{New Estimates of the conversion factor $\eta_{0}$\tablenotemark{a} for Z=Z$_{\odot}$ and $\Phi$(M)$\propto$M$^{-2.35}$}
  \tablehead{\colhead{M$_{up}$}&\colhead{$\eta_{0}$}&\colhead{$\eta_{0}^{\prime}$}&\colhead{$\eta_{0}^{\prime\prime}$}\\
  \colhead{M$_{\odot}$}& \colhead{}& \colhead{}&\colhead{}}
  \startdata
  
 18 &--  &   0.045&  0.091 \\      
   20 &-- &  0.061&  0.139   \\  
    23& -- & 0.093&  0.209   \\            
    25& --  &  0.121  &  0.262   \\     
     28& --  & 0.169 & 0.354   \\    
     30&   0.186 & 0.200  &0.424   \\     
      33 & --& 0.246& 0.544 \\       
    35 &--  &0.278 &0.635   \\      
     40 &--  & 0.359& 0.892  \\             
      45 & 0.393  & 0.445 &1.130  \\     
       50& -- &0.527& 1.347   \\      
       60 & 0.598 &0.674 &1.739   \\

  \enddata
 \tablenotetext{a} {\scriptsize $\eta_{0}$ is taken directly from \cite{vacca94}, we calculate $\eta_{0}^{\prime}$ using the models presented in \cite{vacca94}, and we calculate $\eta_{0}^{\prime\prime}$  based on the model presented in \cite{martins}. In our calculations, we cannot accurately report $\eta_{0}$ for upper masses above 60 M$_{\odot}$ due to the lack of data above this mass in \cite{martins}.}
\end{deluxetable}

\bibliography{hrcrefs}
\bibliographystyle{apj}

\clearpage



\clearpage


\end{document}